\documentclass[preprint,aps,amsmath,amssymb,superscriptaddress,prb,longbibliography]{revtex4-1}   
\usepackage[latin2]{inputenc}
\usepackage[T1]{fontenc}
\usepackage{amsmath,amssymb}
\usepackage{amsfonts}
\usepackage{bm}
\usepackage{dcolumn}
\usepackage{setspace}
\usepackage{graphicx}
\usepackage{color}
\usepackage{multirow}
\usepackage{verbatim}
\usepackage{epstopdf}

\def\bar{\begin{array}}
\def\ear{\end{array}}

\def\s{\sigma}
\def\f{\frac}

\def\R{\mathbf{R}}
\def\r{\mathbf{r}}

\def\nn{\nonumber}

\def\m{\omega^*}

\def\p{\partial}

\def\p{\partial}
\def\a{\alpha}
\def\b{\beta}
\def\g{\gamma}

\def\l{\lambda}

\def\m{\mu}
\def\n{\nu}
\def\r{\rho}
\def\s{\sigma}
\def\t{\tau}
\def\z{\zeta}
\def\e{\epsilon}

\begin{document}

\title{Adiabatic perturbation theory of energy transfer and charge transport in condensed matter}

\author{Ryan Requist} 
\affiliation{Ru\dj er Bo\v{s}kovi\'{c} Institute, Bijeni\v{c}ka cesta 54, 10000 Zagreb, Croatia}
\author{Fran Dabo} 
\affiliation{Department of Physics, Faculty of Science, University of Zagreb, Bijeni\v{c}ka cesta 32, 10000 Zagreb, Croatia}

\date{\today}

\begin{abstract}

Energy and charge transfer in condensed matter systems are often described in terms of quasiparticles.  The order of accuracy of the results in terms of the electron-to-nucleus mass ratio $\e$, an adiabatic small parameter, is usually not known.  To quantify energy transfer with controlled accuracy in $\e$, we use adiabatic perturbation theory to derive a formula for the rate of change of the nuclear kinetic energy.  Applying a sequence of near-identity unitary transformations decouples the electronic and nuclear degrees of freedom to higher and higher order and produces an effective Hamiltonian.  In the special case that the nuclei are treated classically, the energy transfer formula reduces to the usual formula for the rate of work done on a classical particle but the mass of the particle is enhanced by the factor $M_0^{-1}(M_0+\epsilon M_1)$, where $M_1$ is a nonadiabatic correction to the bare nuclear mass tensor $M_0$.  Additional quantum geometric corrections appear at higher orders.  Adiabatic perturbation theory is further used to incorporate, order-by-order in $\e$, nonadiabatic transitions into linear response calculations.  The nonadiabatic frequency-dependent conductivity and Drude weight of an electron-ion system are calculated.
 
\end{abstract}

\maketitle

\section{Introduction}

Non-empirical calculations of condensed matter systems have the advantage of being predictive and free from the uncertainties concerning transferability and applicability that are inherent to empirical calculations.  For a calculation to be non-empirical, it must start from the microscopic Hamiltonian of electrons and nuclei.  Given the complexity of the many-electron many-nucleus problem, it is necessary to invoke several approximations.  Some of the commonly used approximations are uncontrolled, meaning the order of magnitude of the error is not known in terms of a dimensionless small parameter.  Most state-of-the-art calculations of crystalline solids rely on a reference state of harmonic nuclear vibrations and weakly interacting electronic quasiparticles.

The harmonic approximation is the second order truncation of a Hamiltonian that has been expanded in powers of nuclear displacements with respect to a reference configuration.  It is motivated by the fact that nuclei are usually strongly localized.  In molecular systems, Born and Oppenheimer introduced a perturbative approach based on expanding the adiabatic electronic Hamiltonian in powers of scaled nuclear displacements, i.e.~displacements multiplied by $\e^{1/4}$, where $\e=m/M$ is the electron-to-nucleus mass ratio \cite{born1927}.   When the expansion is applied to solids, phonons are readily obtained from the harmonic approximation to the effective nuclear Schr\"odinger equation of the adiabatic electronic ground state.  

The harmonic approximation is not always a valid approximation.  For localized states, the reference configuration is usually chosen to be a point where the adiabatic potential energy surface has a minimum.  This choice is in fact required by the solvability conditions in Born and Oppenheimer's nuclear displacement perturbation theory.  There are several interesting systems where a minimum of the adiabatic potential is not a suitable reference point.  For example, as a consequence of the Jahn-Teller effect \cite{jahn1937}, a high-symmetry point where two or more adiabatic electronic states are degenerate (a point of conical intersection of the adiabatic potential energy surfaces) is surrounded by multiple degenerate potential wells.  The harmonic approximation with respect to the minimum of any well yields a qualitatively incorrect nuclear density, since the true eigenstates are delocalized over multiple wells.  Attempting to correct this by variationally optimizing a superposition of localized states yields eigenstates with incorrect energetic ordering.  To obtain the correct ordering in a method that uses a single adiabatic electronic state, the effective nuclear Schr\"odinger equation must contain an effective vector potential, the Berry-Mead-Truhlar gauge potential \cite{mead1979,berry1984}.  In Jahn-Teller systems, the high-symmetry point is often a more suitable reference point for an expansion of the Hamiltonian, provided that the relevant multiplet of degenerate electronic states can be included in the calculation.  The harmonic approximation is not a valid zeroth order approximation for delocalized states and may be inadequate at high energies and in time-dependent, finite temperature, and nonequilibrium problems.

The harmonic approximation plays a role in several theoretical frameworks that do not rely on the notion of an adiabatic potential energy surface.  A complete set of nonperturbative equations for the electronic and phononic Green's functions and screened interactions, collectively called the Hedin-Baym equations \cite{baym1961,hedin1965,hedin1969}, has been derived within the harmonic approximation \cite{giustino2017}. The phononic Green's function is a nuclear displacement correlation function defined by an equation that resembles the equation defining adiabatic phonons but with a frequency-dependent self-energy in place of the interatomic force constants.  Non\-adiabatic phonons have also been defined using a frequency-dependent dynamical matrix \cite{calandra2010}.

A diagrammatic approach to the equilibrium and nonequilibrium Hedin-Baym equations expands the electronic and phononic self-energies in terms of the interacting electronic Green's function, an interacting $2\times 2$ phononic Green's function describing both nuclear displacements and their conjugate momenta, and screened interactions \cite{stefanucci2023}. It rests on the twin assumptions that the nuclei stay close to their equilibrium positions and that the fluctuations of the electronic density are small.  

The applicability of theoretical methods that do not rely on the adiabatic approximation is not limited to the adiabatic regime.  Yet, to the extent that such methods are independent of the adiabatic approximation, they do not exploit the smallness of the parameter $\e$ and their order of accuracy in the $\e\rightarrow 0$ limit is difficult to determine. 

The harmonic approximation is frequently adopted in condensed matter calculations that use density functional theory \cite{hohenberg1964,kohn1965} for the electronic part of the problem.  Adiabatic phonons, the Kohn-Sham electronic band structure, and electron-phonon interactions can be efficiently calculated with density functional perturbation theory \cite{baroni2001,ponce2016,li2023}.  The reference state consisting of adiabatic phonons and Kohn-Sham electrons has been used to define a zeroth order Hamiltonian and set up a many-body perturbation theory within the harmonic approximation \cite{marini2015}.  Density functional theory yields the electronic density of the Born-Oppenheimer ground state rather than the electronic density of the true ground state of the full system of electrons and nuclei.  Generalizations of density functional theory to the full system of electrons and nuclei incorporate nonadiabatic effects without abandoning the useful concept of an auxiliary system of noninteracting electrons, yet new functional approximations that account for the nonadiabaticity of the electronic state are needed \cite{gidopoulos1998,kreibich2001,butriy2007,requist2016b,li2018,requist2019,fromager2024,cohen2025,wang2025}.

Apart from the harmonic approximation, the second common assumption in condensed matter calculations is that the electronic state 
can be described in terms of weakly interacting quasiparticles.  In the Hedin-Baym equations and similar approaches, electronic quasiparticles interact through a frequency-dependent screened Coulomb interaction \cite{hedin1965,hybertsen1986,onida2002,vanleeuwen2004,giustino2017,harkonen2020,stefanucci2023}.  The screened Coulomb interaction is shorter ranged and generally weaker than the bare Coulomb interaction.  However, in some materials, such as Mott insulators, the screened interaction is still too strong for the electronic state to be adequately described in terms of weakly interacting quasiparticles.  Since there exists a great variety of electronic structure methods, and the choice of the method often depends on the problem at hand, it is desirable, when formulating adiabatic approximations for systems of electrons and nuclei, to maintain the broadest possible applicability by designing a theory that makes no assumptions about the method used to treat the electronic part of the problem.

Our focus is on problems in condensed matter physics where the harmonic approximation does not provide an acceptable zeroth order solution.  This includes systems with multiple degenerate minima and time-dependent, finite temperature, and nonequilibrium problems in which the nuclear state becomes delocalized or the nuclei undergo large amplitude motion or tunneling.  In the microscopic Hamiltonian, the adiabatic small parameter $\e$ multiplies the nuclear kinetic energy operator but does not appear anywhere else.  Hence, adiabatic approximations for the $\e\rightarrow 0$ limit do not depend on the validity of the harmonic approximation. This motivated the development of a perturbative approach, called adiabatic perturbation theory, in which the nuclear kinetic energy operator is treated perturbatively in the $\e\rightarrow 0$ limit without assuming an expansion in nuclear displacements \cite{requist2025}.  

Adiabatic perturbation theory employs a sequence of increasingly refined unitary transformations to transform to a sequence of higher order adiabatic representations in which the microscopic Hamiltonian is more nearly diagonal with respect to the electronic states.  Neglecting the off-diagonal elements of the Hamiltonian in the $n$th order adiabatic representation yields an effective Hamiltonian.  As this strategy does not depend on an expansion in nuclear displacements, it is not limited to localized states.  Since all physical results are invariant to unitary changes of representation, any of the higher order adiabatic representations can serve as an equally valid alternative starting point for the theoretical description of the system.  Using an effective Hamiltonian as a starting point does not preclude the use of the harmonic approximation or a quasiparticle-based approximation in subsequent stages.

Here, we make two developments.  First, we extend adiabatic perturbation theory to two-component systems that vary slowly in time and derive the lowest order nonadiabatic correction to the rate of change of the nuclear kinetic energy.  Second, within linear response theory we show how the generators of the unitary transformations to higher order adiabatic representations give simple and systematic formulas---to any order in $\e$---for the nonadiabatic corrections to linear response functions. 

The effects of ionic motion on electronic properties are usually accounted for through the electron-phonon interaction.  In electron-phonon Hamiltonians, the nuclear kinetic energy operator is typically modeled as $\sum_{q\l} P_{q\l}^{\dag}P_{q\l}/2M_{q\l}$, with $P_{q\l}=i \sqrt{\hbar\omega_{q\l} M_{q\l}/2} (a_{q\l}^{\dag}-a_{-q\l})$ given in terms of the phonon creation and annihilation operators.  This model nuclear kinetic energy operator is not equivalent to the nuclear Laplacian in the microscopic Hamiltonian \cite{vanleeuwen2004}.  To explicitly demonstrate that adiabatic perturbation theory enables calculations without an auxiliary electron-phonon Hamiltonian, we introduce a continuous one-dimensional electron-ion system whose Hamiltonian retains the differential Laplacians of the electronic and nuclear kinetic energy operators.  We do not impose a periodic lattice structure from the outset.  The electrons are fully correlated, as no approximations are made in the electronic part of the problem.  We perform numerical calculations of the Drude weight and frequency-dependent electrical conductivity.  

The paper is organized as follows.  Adiabatic perturbation theory and the notion of higher order adiabatic representations are introduced in Sec.~\ref{sec:APT}.  In Sec.~\ref{sec:TD}, the theory is extended to 
slowly varying time-dependent systems and a nonadiabatic formula for the rate of change of the nuclear kinetic energy is derived.  Linear response theory of electron-ion systems including the nonadiabatic corrections due to ionic motion is formulated in Sec.~\ref{sec:LR}.  A continuous one-dimensional model system is introduced in Sec.~\ref{sec:model}.  Numerical calculations of the nonadiabatic electrical conductivity and Drude weight in this model are reported in Sec.~\ref{sec:conductivity}.   The Discussion section contains a perspective on how adiabatic perturbation theory differs from conventional approaches.  The Conclusions summarize our findings and give an outlook on future research.

\section{Adiabatic perturbation theory \label{sec:APT}}

We begin by introducing an adiabatic small parameter $\e$ into the Hamiltonian of a system of electrons and nuclei.  The form of the resulting Schr\"odinger equation shows that it is a singularly perturbed differential equation in the adiabatic limit $\e\rightarrow 0$.  Due to the limitations of regular perturbation theory in treating singularly perturbed problems, we introduce an alternative method.  It belongs to a class of methods known as singular perturbation theory or adiabatic perturbation theory.

The nonrelativistic Hamiltonian of electrons and nuclei in laboratory frame coordinates $\mathbf{r}_i$ and $\R_i$ and in atomic units is
\begin{align}
\mathcal{H} &= -\sum_{i} \f{1}{2M_i}\nabla_{\R_i}^2 - \sum_i \f{1}{2} \nabla_{\mathbf{r}_i}^2 + \sum_{i<j} \f{1}{|\mathbf{r}_i-\mathbf{r}_j|} - \sum_{i,j} \f{Z_j}{|\mathbf{r}_i-\mathbf{R}_j|} + \sum_{i< j} \f{Z_i Z_j}{|\mathbf{R}_i-\mathbf{R}_j|} {.}
\label{eq:H}
\end{align}
We want to make use of the fact that the nuclear masses $M_i$ are much larger than the electron mass to develop adiabatic approximations and quantify their error.  For this purpose, we introduce a small parameter $\e$ and scale all of the nuclear masses according to
\begin{align}
M_i \rightarrow \e^{-1} M_i {.}
\label{eq:scaling}
\end{align}
Applying this scaling to Eq.~(\ref{eq:H}) introduces a factor of $\e$ multiplying the nuclear kinetic energy operator.  We will see that this gives us a way of deriving adiabatic approximations and characterizing their accuracy in terms of a single small parameter.

It is desirable to express the nuclear kinetic energy operator in a coordinate invariant form with generalized coordinates $\{x^{\m}\}$ instead of the laboratory frame coordinates $\{\R_i\}$.  This is useful for handling internal coordinates, for which there is no unique choice.  Internal coordinates arise after splitting off the center-of-mass coordinate (as we do in Sec.~\ref{sec:model}).  The coordinate invariant expression for the nuclear kinetic energy operator in terms of the Laplace-Beltrami operator is 
\begin{align}
T = -\frac{1}{2} \frac{1}{\sqrt{g_0}} \partial_\mu \sqrt{g_0} g^{\mu\nu}_0 \partial_\nu {,}
\label{eq:T}
\end{align}
where $\p_{\m}=\p/\p x^{\m}$, $g^{\mu\nu}_0$ is the inverse of the bare mass tensor $g_{0\mu\nu}$, and $g_0=\mathrm{det}(g_{0\m\n})$.  The full Hamiltonian is 
\begin{align}
    \mathcal{H} = \epsilon T + H,
    \label{eq:H:static}
\end{align}
where $H=H(x)$ is the Born-Oppenheimer Hamiltonian, i.e., the last four terms of Eq.~(\ref{eq:H}) expressed in terms of the $x$ coordinates. 

We can now write the Schr\"odinger equation in the standard form 
\begin{align}
( \epsilon T + H)|\psi\rangle = \Omega |\psi\rangle {.}
\label{eq:schroedinger}
\end{align}
We want to find approximate solutions in the $\e\rightarrow 0$ limit.  Equation (\ref{eq:schroedinger}) is a singularly perturbed differential equation in the $\e\rightarrow 0$ limit.  A singularly perturbed differential equation is one whose character changes qualitatively when the perturbation is set to zero.  Setting $\e=0$ in Eq.~(\ref{eq:schroedinger}) changes it from a differential equation in which both the electronic and nuclear coordinates are independent variables to a differential equation in which the nuclear coordinates $x$ are merely parameters, 
\begin{align}
    H(x) |n(x)\rangle = E_n(x) |n(x)\rangle {,}
    \label{eq:adiabatic}
\end{align}
where $n$ denotes the quantum numbers of the eigenstate.  This is an independent eigenvalue problem for each value of $x$.  The adiabatic eigenstates $|n\rangle=|n(x)\rangle$, being elements of the electronic Hilbert space, are not valid solutions of the full Schr\"odinger equation in Eq.~(\ref{eq:schroedinger}).  In particular, they are not normalizable in $x$-space and do not obey the boundary conditions.

At this point, it is clear we have a non-standard problem.  Regular perturbation theory assumes that the unperturbed problem is solvable, i.e.~that it provides zeroth order eigenstates that obey the boundary conditions.  Yet the equation obtained by setting $\e=0$ in Eq.~(\ref{eq:schroedinger}) does not provide such zeroth order solutions.  The unsuitability of regular perturbation theory is discussed further in Sec.~\ref{sec:discussion}.

Before introducing adiabatic perturbation theory for condensed matter systems, we briefly describe how $\e$ can be used to specify the accuracy of an adiabatic approximation.  Solving the time-independent Schr\"odinger equation in Eq.~(\ref{eq:schroedinger}) within an adiabatic approximation produces an $\e$-dependent wave function $|\psi^{approx}\rangle$.  Using $|\psi^{approx}\rangle$ to evaluate the expectation value of an observable $O$ yields an $\e$-dependent expression $\langle O\rangle^{approx}$.  The exact expectation value $\langle O\rangle^{exact}$ is also $\e$ dependent.  The order of accuracy of an adiabatic approximation is determined by how the error scales with $\e$ in the $\e\rightarrow 0$ limit.  The result of an adiabatic approximation will be said to be accurate to $n$th order in $\e$ if
\begin{align}
\lim_{\e\rightarrow 0} \e^{-n} \left|\langle O\rangle^{approx} - \langle O\rangle^{exact}\right| = 0 {.}
\end{align}
This implies that the error vanishes faster than $\e^n$.  To predict the physical value of an observable within a given adiabatic approximation, we simply set $\e$ to unity in $\langle O\rangle^{approx}$.

We now introduce adiabatic perturbation theory.  An electron-ion state can be expressed in the adiabatic representation as
\begin{align}
    |\psi(x)\rangle = \sum_n \zeta_n(x)|n(x)\rangle {.}
    \label{eq:psi}
\end{align}
This is a mixed notation, where $|\psi(x)\rangle$ stands for $\langle x|\psi\rangle$, which is a partial projection of $|\psi\rangle$, namely the projection to the ionic position representation.  The state in the full electron-ion position representation is $\psi(r,x) = \langle r|\psi(x)\rangle$.  There is an ionic wave function $\z_n=\zeta_n(x)$ associated with each adiabatic electronic eigenstate $|n\rangle$; we will use the terms ``ionic'' and ``nuclear'' interchangeably.  Substituting Eq.~(\ref{eq:psi}) into the Schr\"odinger equation and projecting onto $|l\rangle$ gives
\begin{align}
  -\frac{\epsilon}{2} \sum_{mn} \f{1}{\sqrt{g_0}} (\partial_\mu\delta_{lm} - iA_{\mu,lm})\sqrt{g_0} g^{\mu\nu}_0 (\partial_\nu\delta_{mn} - iA_{\nu,mn})\zeta_{n}+ E_l(x)\zeta_{l} = \Omega \zeta_{l}{.}&
    \label{eq:adiabatic:0}
\end{align}
The operator in parentheses is the $U(N)$-gauge covariant derivative, $D_{\mu,ln} = \partial_\mu \delta_{ln} - iA_{\mu,ln}$, where $N$ is the dimension of the electronic Hilbert space, and $A_{\mu,ln}$ is a $U(N)$ gauge potential, which is also called the nonadiabatic coupling.  

Equation (\ref{eq:adiabatic:0}) is the gauge- and coordinate-invariant electron-ion Schr\"odinger equation in the adiabatic representation.  In this representation, an electron-ion state $|\psi(x)\rangle$ is uniquely specified by a column vector $[\zeta(x)]=(\z_1(x),\z_2(x),\ldots)^T$ whose elements are the ionic wave functions.  There is a one-to-one mapping between states $|\psi(x)\rangle$ and column vectors $[\zeta(x)]$.   Equation~(\ref{eq:adiabatic:0}) can be written as a matrix equation $[\mathcal{H}(x)] [\zeta(x)]  = \Omega [\zeta(x)]$ in which the elements of the Hamiltonian matrix $[\mathcal{H}(x)]$ contain differential operators, i.e.
\begin{align}
\mathcal{H}_{ln}(x) = -\frac{\epsilon}{2} \sum_{m} \f{1}{\sqrt{g_0}} (\partial_\mu\delta_{lm} - iA_{\mu,lm}) \sqrt{g_0} g^{\mu\nu}_0 (\partial_\nu\delta_{mn} - iA_{\nu,mn})+ E_l(x)\delta_{ln} {.}
    \label{eq:H0}
\end{align}
From now on, we refer to the usual adiabatic representation, used in Eqs.~(\ref{eq:psi})-(\ref{eq:H0}), as the {\it zeroth order adiabatic representation}.  Quantities defined with respect to this representation will be labeled with a superscript $(0)$, e.g.~the electron-ion Hamiltonian will be denoted as $\mathcal{H}^{(0)}(x)$, although this label will sometimes be suppressed.  The first term of Eq.~(\ref{eq:H0}) is the matrix element of the ionic kinetic energy operator 
\begin{align}
T_{ln}^{(0)} &= -\frac{1}{2} \sum_{m} \f{1}{\sqrt{g_0}} (\partial_\mu\delta_{lm} - iA_{\mu,lm}) \sqrt{g_0} g^{\mu\nu}_0 (\partial_\nu\delta_{mn} - iA_{\nu,mn}) {.}
\label{eq:Tln}
\end{align}

So far no approximation has been made.  But if we neglect the coupling between different electronic states due to the off-diagonal elements of $T_{ln}^{(0)}$, we obtain a set of decoupled effective Schr\"odinger equations
\begin{align}
\e T^{{\rm eff}(0)}_n \zeta_{n\g} + E_n \zeta_{n\g} + \epsilon V_{geo,n} \zeta_{n\g}  = \Omega_{n\g} \zeta_{n\g} & {,}
\label{eq: first order adiabatic Schrodinger}
\end{align}
which contain the {\it zeroth order effective ionic kinetic energy operator}
\begin{align}
T^{{\rm eff}(0)}_n &= -\f{1}{2} \f{1}{\sqrt{g_0}} \big( \p_{\m} - i A_{\m,nn} \big) \sqrt{g_0} g_0^{\m\n} \big( \p_{\n} - i A_{\n,nn} \big) {}
\label{eq:Tkin:0}
\end{align}
and the geometric scalar potential \cite{jackiw1988,berry1989}
\begin{align}
V_{geo,n}(x) = \frac{1}{2} g^{\mu\nu}_0\sum_{k\neq n} A_{\mu,nk}(x) A_{\nu,kn}(x)
\label{eq:Vgeo}
\end{align}
that accounts for virtual transitions from state $n$ to state $k$ and back to state $n$. The gauge potential $A_{\mu,nn}=A_{\mu,nn}(x)$ is the Berry-Mead-Truhlar gauge potential \cite{mead1979,berry1984}.  

Equation (\ref{eq: first order adiabatic Schrodinger}) is manifestly gauge- and coordinate-invariant and this form of Schr\"odinger equation results from partitioning $T_{nn}^{(0)}$ into two gauge- and coordinate-invariant terms, i.e.
\begin{align}
T_{nn}^{(0)} &= T^{{\rm eff}(0)}_n + V_{geo,n} {.}
\label{eq:Tnn:partitioning}
\end{align} 
where all of the differential operators are contained in the first term.  Under the gauge transformation $\z_{n\g}(x) \rightarrow e^{i\lambda_n(x)}\z_{n\g}(x)$; $|n(x)\rangle \rightarrow e^{-i\lambda_n(x)}|n(x)\rangle$, the gauge potential transforms as $A_{\m,nn} \rightarrow A_{\m,nn} + \p_{\m} \lambda_n$.  The index $\g$ denotes the ionic quantum numbers of the wave function $\z_{n\g}=\z_{n\g}(x)$ associated to the $n$th electronic state.  Let $[\zeta^{{\rm eff}(0)}(x)] = (\ldots, 0,\z_{n\g}(x),0,\ldots)^T$ be the column vector whose only nonzero element, the one in the $n$th position, is the ionic eigenfunction $\z_{n\g}(x)$ from Eq.~(\ref{eq: first order adiabatic Schrodinger}). The column vector $[\zeta^{{\rm eff}(0)}(x)]$ maps to a state $|\psi^{{\rm eff}(0)}(x)\rangle = \z_{n\g}(x)|n(x)\rangle$ that approximates one of the exact electron-ion eigenstates.  The eigenenergy $\Omega_{n\g}$ approximates the total energy of the electron-ion system.   A similar splitting of the nuclear kinetic energy into two gauge-invariant terms has been made within the context of the exact factorization method and identities have been derived for the rate of change of each term \cite{li2022,requist2022}.  Hereafter, we will suppress the $n$ subscripts on $T^{{\rm eff}(0)}_{n}$, $V_{geo,n}$, $A_{\m,nn}$, and related quantities, except when it might cause confusion.

Equation~(\ref{eq: first order adiabatic Schrodinger}) contains an effective Hamiltonian for the electron-ion system, which will be called the {\it zeroth order effective Hamiltonian} and denoted as $\mathcal{H}^{{\rm eff}(0)}(x)$.  $\mathcal{H}^{{\rm eff}(0)}(x)$ is obtained by setting the off-diagonal elements of $\mathcal{H}^{(0)}(x)$ to zero.  Its diagonal elements are
\begin{align}
\mathcal{H}^{{\rm eff}(0)}_{nn}(x) = \e T^{{\rm eff}(0)}_n + E_n(x) + \epsilon V_{geo,n}(x) {.}
\label{eq:Heff:0}
\end{align}

The idea behind adiabatic perturbation theory is to perform a sequence of unitary transformations to successively higher order adiabatic representations in which the Hamiltonian becomes more and more nearly diagonal with respect to the adiabatic electronic states.  That is, the transformations should make the coupling between electronic states of higher and higher order in $\e$.  The first transformation, which takes the system from the zeroth order to the first order adiabatic representation, removes the coupling to order $\e$.  The transformation is accomplished by the unitary operator
\begin{align}
U_1 = e^{i\e G_1} {.}
\end{align}
The Hermitian generator $G_1$ is chosen such that the off-diagonal elements of the transformed Hamiltonian matrix $[\mathcal{H}^{(1)}]=[U_1^{\dag}] [\mathcal{H}^{(0)}][U_1]$ vanish to order $\e$.  Using a Baker-Campbell-Hausdorff-type formula to expand $\mathcal{H}^{(1)}$ yields
\begin{align}
\mathcal{H}^{(1)} &= H + \e T - i\epsilon[G_1,H] - i\epsilon^2 [G_1,T] - \f{1}{2}\e^2 [G_1,[G_1,H]] + \mathcal{O}(\epsilon^3) {.}
\label{eq:H(1)}
\end{align}
The order $\e$ off-diagonal elements of the Hamiltonian matrix will vanish if $G_1$ satisfies the operator equation
\begin{align}
T_{ln} - i [G_1,H]_{ln} = 0 \quad \textrm{for} \quad l\neq n {,} 
\end{align} 
which determines the off-diagonal elements of $G_1$ to be \cite{requist2025}
\begin{align}
G_{1,ln} &= -\frac{i}{2}g^{\mu\nu}_0 \sum_{m} D_{\mu,lm} \frac{1}{E_l-E_n} D_{\nu,mn} - g^{\mu\nu}_0\frac{1}{2} \frac{\partial_\mu(E_l+E_n)}{(E_l-E_n)^2} A_{\nu,ln} \nn \\
&= \f{i}{E_l-E_n} T_{ln} - i g_0^{\m\n} \f{\p_{\m} E_n}{(E_l-E_n)^2}(-iA_{\n,ln}) {.}
\label{eq:G1}
\end{align}
The diagonal elements of $G_1$ remain undetermined.  We choose them to be zero, which is a gauge choice.

Transformations to higher order adiabatic representations can be made by applying a sequence of increasingly refined near-identity unitary operators of the form
\begin{align}
U_k = e^{i\e^k G_k} {.}
\label{eq:Uk}
\end{align}
A state $|\psi^{(0)}\rangle$ in the zeroth order adiabatic representation transforms to the state 
\begin{align}
|\psi^{(p)}\rangle  = U_p^{\dag} \ldots U_1^{\dag} |\psi^{(0)}\rangle {}
\end{align}
in the $p$th order adiabatic representation, while an operator $O^{(0)}$ transforms to
\begin{align}
O^{(p)} = U_p^{\dag} \dots U_1^{\dag} O^{(0)} U_1\dots U_p {.}
\end{align}

The higher order adiabatic representations form a hierarchy.  To pass from level $p-1$ to level $p$, the generator $G_p$ is chosen to make the off-diagonal elements in the $p$th order Hamiltonian matrix $[\mathcal{H}^{(p)}]$ vanish to order $\e^p$.  The $p$th order effective Hamiltonian $\mathcal{H}^{{\rm eff}(p)}$ is obtained by neglecting the off-diagonal elements of $[\mathcal{H}^{(p)}]$ and truncating its diagonal elements to order $\e^{p+1}$, i.e.
\begin{align}
\mathcal{H}^{{\rm eff}(p)}_{ln}(x) = \left\{ \begin{array}{ll} \mathcal{H}^{(p)}_{ln}(x)\big|_{\textrm{truncated to order}\;\e^{p+1}} & l=n \\
0 & l\neq n \end{array} \right. {.}
\end{align}
Here, the order $p$ of the effective Hamiltonian denotes the order of the higher order adiabatic representation in which it was obtained.  This is different than the convention used in Ref.~\onlinecite{requist2025}, where the effective Hamiltonians were labeled according to their order of accuracy.  Despite the different labeling convention, the Hamiltonians are exactly the same.  Specifically, a Hamiltonian labeled here as $\mathcal{H}^{{\rm eff}(p)}$ is the same as a Hamiltonian labeled $\mathcal{H}^{{\rm eff}(p+1)}$ in Ref.~\onlinecite{requist2025}. 

We illustrate this procedure for $p=1$, using the generator $G_1$ in Eq.~(\ref{eq:G1}). 
Substituting $G_1$ into $\mathcal{H}^{(1)}$ in Eq.~(\ref{eq:H(1)}), neglecting the remaining (order $\e^2$) off-diagonal coupling, and truncating the diagonal elements to order $\e^2$, we obtain the first order effective Hamiltonian $\mathcal{H}^{{\rm eff}(1)}$ with the diagonal elements
\begin{align}
\mathcal{H}^{{\rm eff}(1)}_{nn} = -\frac{\epsilon}{2} \f{1}{\sqrt{g_0}}(\partial_\mu -iA_{\mu}^{(1)*}) \sqrt{g_0} h^{(1)\mu\nu}_0 (\partial_\nu -iA_{\nu}^{(1)}) + E_n+ \epsilon V_{geo} + \e^2 V_{2,geo} {,}
\label{eq:Heff:1}
\end{align}
where $h^{(1)}_{\m\n} = g_{0\m\n} +\e h_{1\m\n}$ is a Hermitian mass tensor, $A_{\m}^{(1)}= A_{\m}+\e A_{1\m}$, $A_{1\m}$ is a complex-valued vector potential, and $V_{2,geo}$ is a second order correction to the potential.  
The formulas defining $h_1$, $A_{1\m}$, and $V_{2,geo}$ are 
\begin{align}
h_{1\m\n} &= -2 \langle D_{\m} n | (E_n-H)^{-1} | D_{\n} n\rangle \nn \\
A_{1\n} &= \langle D_{\n} n |(E_n-H)^{-1} |\nabla^2 n\rangle - \langle D_{\n} n |(E_n-H)^{-2}| D_{\r} n\rangle g_0^{\r\s} \p_{\s} E_n \nn \\
V_{2,geo} &= \f{1}{4} \langle \nabla^2 n | (E_n-H)^{-1} | \nabla^2 n \rangle -\f{1}{2}  g_0^{\m\n} \p_{\n} E_n \mathrm{Re} \langle D_{\m} n |(E_n-H)^{-2} | \nabla^2 n \rangle {}
\label{eq:nonadiabatic quantities}
\end{align}
with
\begin{align}
|\nabla^2 n\rangle &\equiv g_0^{\r\s} (1-|n\rangle\langle n|) |D_{\r} D_{\s} n\rangle - g_0^{\r\s} \Gamma_{0\r\s}^{\t} |D_{\t} n\rangle {,}
\label{eq:nablasq}
\end{align}
where $\Gamma_{0\m\n}^{\l}$ are the components of the Levi-Civita connection $\nabla$ compatible with $g_{0\m\n}$.  The first term in Eq.~(\ref{eq:Heff:1}) is a Hermitian operator, as it must be since it was obtained by the unitary transformation and consistent truncation of the microscopic Hamiltonian. 

Adiabatic approximations based on unitary transformations to higher order adiabatic representations have been developed for slowly driven time-dependent quantum systems \cite{berry1987,requist2010,requist2023} and electron-ion systems \cite{littlejohn1993,weigert1993,teufel2003,matyus2019,littlejohn2024,chatzistavrakidis2026}.  Refs.~\onlinecite{littlejohn1993,weigert1993,teufel2003,matyus2019,littlejohn2024} obtained effective Hamiltonians for large amplitude motion in molecules when the nuclear kinetic energy is of order 1.  The effective Hamiltonians are therefore different than the ones discussed in this section.  In condensed matter systems, the nuclear kinetic energy is usually of order $\e^{1/2}$, as is the case in the model system we study in Sec.~\ref{sec:model}.  However, there are some problems which have simultaneously some nuclear degrees of freedom with kinetic energy of order $\e^{1/2}$ and others with kinetic energy of order 1.  For example, in the scattering of a molecule on a surface, the center-of-mass coordinate of the molecule undergoes large amplitude motion and can be treated as having kinetic energy of order 1, while the atoms of the substrate can usually be treated as having kinetic energy of order $\e^{1/2}$, unless they are strongly perturbed following the collision with the incoming molecule.  The adiabatic perturbation theory discussed in this section treats both types of nuclear degrees of freedom simultaneously with high accuracy.

In this section, adiabatic perturbation theory was used to diagonalize an electron-ion Hamiltonian.  The theory can be straightforwardly adapted to block diagonalization by choosing the generators to remove the coupling between two different sets of states to a given order in $\e$.  Block diagonalization has been applied to molecules in Refs.~\onlinecite{matyus2019,littlejohn2024}.  Block diagonalization is useful when the nonadiabatic coupling among a set of states is so strong that the system cannot be adequately approximated as occupying predominantly a single state up to small nonadiabatic corrections.

The fundamental ingredients in adiabatic perturbation theory are the $x$-dependent adiabatic electronic eigenstates and eigenvalues defined by Eq.~(\ref{eq:adiabatic}).  Adiabatic perturbation theory can be combined with any electronic structure method that can approximate these quantities with sufficient accuracy.

\section{Energy transfer in the adiabatic regime \label{sec:TD}}

The amount of energy exchanged between electrons and ions, a single time-dependent scalar that aggregates information from many degrees of freedom, is a useful quantity for tracking physical processes.  Electrons excited by external electromagnetic fields transfer some of their excess energy to lattice vibrations.  Electronic currents in wires excite ionic vibrations in a process known as Joule heating \cite{diventra2002,horsfield2004}. Energy can also flow in the opposite direction, from atomic and molecular motion to electronic degrees of freedom.  For example, the vibrations of a diatomic molecule adsorbed on a metal surface are damped by the creation of electron-hole pair excitations \cite{rittmeyer2015,hopjan2018,bombin2023}. 

Energy transfer in condensed matter systems is usually described in terms of quasiparticle scattering mediated by the electron-phonon coupling \cite{allen1987,waldecker2016}.  Here, we follow a different path and describe energy transfer in higher order adiabatic representations.  The approximate separation of electronic and ionic variables that is achieved by the transformation to a higher order adiabatic representation simplifies the problem to an effective Schr\"odinger equation for the ionic degrees of freedom.  

We apply our theory to two dynamical problems.  In the first, we consider the dynamics of an electron-ion system that starts in a nonstationary state, such as the state that is created when a system undergoes a vertical excitation.  A vertical excitation is an idealized excitation that lifts the electrons from their ground state to an excited state but preserves the shape of the ionic wave function.  Adiabatic perturbation theory is used to derive a formula for the rate of change of the ionic kinetic energy.  In the second, we consider an electron-ion system that is slowly driven by external fields and derive an effective Hamiltonian that can be used to quantify energy transfer.

\subsection{Energy transfer during slow ionic motion \label{ssec:slow-ions}}

In the first dynamical scenario, we consider an electron-ion system that starts in a nonstationary initial state following a vertical excitation and subsequently evolves according to the time-dependent Schr\"odinger equation 
\begin{align}
i \f{d}{d t} |\psi(x,t)\rangle = \mathcal{H}(x) |\psi(x,t)\rangle 
\label{eq:td:schroedinger:original}
\end{align}
with the time-independent Hamiltonian $\mathcal{H}(x)=\e T+H(x)$.  Since the Hamiltonian is time-independent, the electron-ion state will be represented in the same adiabatic representation as in Sec.~\ref{sec:APT}, i.e.
\begin{align}
|\psi(x,t)\rangle = \sum_n \zeta_n(x,t) |n(x)\rangle {.}
\label{eq:psi:td:A}
\end{align}
Substituting Eq.~(\ref{eq:psi:td:A}) into Eq.~(\ref{eq:td:schroedinger:original}) and projecting onto $|l\rangle$, we obtain the time-dependent Schr\"odinger equation in the zeroth order adiabatic representation
\begin{align}
i \p_t \zeta_{l} =  -\frac{\epsilon}{2} \sum_{mn} \f{1}{\sqrt{g_0}} (\partial_\mu\delta_{lm} - iA_{\mu,lm})\sqrt{g_0} g^{\mu\nu}_0 (\partial_\nu\delta_{mn} - iA_{\nu,mn})\zeta_{n}+ E_l(x)\zeta_{l} {.}&
    \label{eq:adiabatic:0:td:A}
\end{align}
We will derive the rate of change of the ionic kinetic energy in the zeroth and first order adiabatic approximations.  The $p$th order adiabatic approximation, for any integer $p$, is formulated in the $p$th order adiabatic representation, and in this representation we choose the initial state to be
\begin{align}
|\psi^{(p)}(x)\rangle = \z_{{\rm gs}}^{(p)}(x) |n(x)\rangle {,}
\label{eq:initial:condition}
\end{align}
where $n$ are the quantum numbers of a given excited state.  The ionic wave function $\z_{{\rm gs}}^{(p)}(x)$ is taken to be the ground state wave function on the ground state adiabatic potential energy surface, and we assume it to be localized near a single point $x=x_0$.  It follows that $\z_{{\rm gs}}^{(p)}(x)$ has a width of order $\e^{1/4}$.  In the notation of Sec.~\ref{sec:APT}, the initial state $|\psi^{(p)}(x)\rangle$ maps to the column vector $[\zeta^{(p)}(x)] = (\ldots, 0,\z_{{\rm gs}}^{(p)}(x),0,\ldots)^T$.  We further assume that $x_0$ is not an extremum of the excited state adiabatic potential $E_n(x)$.  

In the zeroth order adiabatic approximation, the off-diagonal elements of the Hamiltonian in Eq.~(\ref{eq:adiabatic:0:td:A}) are neglected and $|\psi^{(0)}(x,t)\rangle$ maps to $[\zeta^{(0)}(x,t)] = (\ldots, 0,\z_{n}^{(0)}(x,t),0,\ldots)^T$, where the wave function $\zeta_{n}^{(0)}(x,t)$ evolves according to the zeroth order time-dependent effective Schr\"odinger equation
\begin{align}
i \p_{t} \zeta_{n}^{(0)} = \e T^{{\rm eff}(0)}\zeta_{n}^{(0)} + (E_n+\epsilon V_{geo}) \zeta_{n}^{(0)} {}
\label{eq: 0th order TDSE}
\end{align}
with the same effective Hamiltonian $\mathcal{H}^{{\rm eff}(0)}_{nn}$ as in the static case [cf.~Eq.~(\ref{eq:Heff:0})].  As the initial state $\z_{{\rm gs}}^{(p)}(x)$ is not an eigenstate of $\mathcal{H}^{{\rm eff}(0)}_{nn}$, the ions have nontrivial dynamics and there is energy transfer between electrons and ions. 

The ionic kinetic energy can be evaluated order-by-order within adiabatic perturbation theory.  In the zeroth order adiabatic representation, 
\begin{align}
\langle T \rangle &= \langle \psi^{(0)} | T^{(0)} | \psi^{(0)} \rangle \nn \\
&= \sum_{ln} \int dx \sqrt{g_0} \z_l^{(0)*} T_{ln}^{(0)} \z_n^{(0)} \nn \\
&= \int dx \sqrt{g_0} \z_n^{(0)*} T_{nn}^{(0)} \z_n^{(0)} + \textrm{higher order terms} {.}
\label{eq:T:expectation value}
\end{align}
In the last line, we have kept only the contribution from the $n$th electronic state in accordance with our choice of initial condition and the zeroth order adiabatic decoupling.  

The hierarchy of effective ionic Schr\"odinger equations in adiabatic perturbation theory provides us with a way to calculate higher order corrections to the ionic motion in the classical limit.  The initial condition that the electrons occupy a single excited adiabatic eigenstate, corresponding to an idealized vertical excitation, coupled with the assumption that $x_0$, the point at which the initial ionic wave function is localized, is not an extremum of $E_n$, implies that the ions have an energy of order 1 with respect to $\hbar$.  Therefore, they move a distance of order 1 on the potential energy surface, and the action for their motion is large with respect to $\hbar$.  The ions are in a semiclassical regime \cite{mead2006,littlejohn2024}.  To study the motion of the ions in this regime, we reintroduce the factors of $\hbar$ associated with $d/dt$ and $T$ and analyze the $\hbar\rightarrow 0$ limit of the time-dependent Schr\"odinger equation
\begin{align}
i \hbar \p_{t} \zeta_{n}^{(0)} = -\frac{\hbar^2}{2} \f{1}{\sqrt{g_0}} (\partial_\mu -iA_{\mu}) \sqrt{g_0} M^{\mu\nu}_0  (\partial_\nu -iA_{\nu})\zeta_{n}^{(0)} + (E_n+\hbar^2 V_{geo,sc}) \zeta_{n}^{(0)} {,}
\label{eq: 0th order TDSE: hbar}
\end{align}
where we have introduced the mass tensor $M_{0\m\n}=\e^{-1} g_{0\m\n}$ and the geometric potential
\begin{align}
V_{geo,sc} &= \frac{1}{2} M^{\mu\nu}_0\sum_{k\neq n} A_{\mu,nk} A_{\nu,kn} \nn \\
&= \f{1}{2} M^{\mu\nu}_0 \langle D_{\m} n | D_{\n} n\rangle {}
\label{eq:Vgeo,sc}
\end{align}
which only differs from $V_{geo}$ by a factor of $\e$.  Here, $D_{\m}=\p_{\m}+iA_{\m}$ is the gauge-covariant derivative acting on $|n\rangle$.  We introduce the parametrization
\begin{align}
\z_{n}^{(0)}(x,t) = \sqrt{\rho(x,t)} e^{i\hbar^{-1}S(x,t)} {,} 
\label{eq:zeta:RS}
\end{align}
where $\rho=\rho(x,t)$ and $S=S(x,t)$ are assumed to have the expansions
\begin{align}
\rho(x,t) &= \rho_0(x,t)+\hbar \rho_1(x,t)+\hbar^2 \rho_2(x,t)+\cdots \nn \\
S(x,t) &= S_0(x,t)+\hbar S_1(x,t)+\hbar^2 S_2(x,t)+\cdots {.}
\label{eq:RS}
\end{align}
Substituting $\z_n^{(0)}= \sqrt{\rho} e^{i\hbar^{-1}S}$ into Eq.~(\ref{eq: 0th order TDSE: hbar}) and dividing by $\zeta_{n}^{(0)}$ yields
\begin{align}
\p_t S +\f{1}{2} &M_0^{\m\n} (\p_{\m} S-\hbar A_{\m}) (\p_{\n} S-\hbar A_{\n}) + E_n + \hbar^2 V_{geo,sc}  \nn\\
&= i\hbar \f{1}{\sqrt{\rho}} \p_t \sqrt{\rho} + i\hbar M_0^{\m\n} (\p_{\m} \ln \sqrt{\rho}) (\p_{\n} S-\hbar A_{\n}) +i \f{\hbar}{2} \f{1}{\sqrt{g_0}} \p_{\m} \big[ \sqrt{g_0} M_0^{\m\n} (\p_{\n} S-\hbar A_{\n}) \big] \nn \\
&\quad + \f{\hbar^2}{2} \f{1}{\sqrt{\rho}} \f{1}{\sqrt{g_0}} \p_{\m} \big[ \sqrt{g_0} M_0^{\m\n} \p_{\n} \sqrt{\rho} \big]  {.}
\label{eq:schroedinger:RS}
\end{align}
The real part gives 
\begin{align}
\p_t S &+\f{1}{2} M_0^{\m\n} (\p_{\m} S-\hbar A_{\m}) (\p_{\n} S-\hbar A_{\n}) + E_n + \hbar^2 V_{geo,sc} - \f{\hbar^2}{2} \f{1}{\sqrt{\rho}} \f{1}{\sqrt{g_0}} \p_{\m} \big( \sqrt{g_0} M_0^{\m\n} \p_{\n} \sqrt{\rho} \big) = 0 {.}
\label{eq:schroedinger:S}
\end{align}
Neglecting the order $\hbar^2$ terms gives the Hamilton-Jacobi equation with corrections from the gauge potential $A_{\n}$.  The imaginary part of Eq.~(\ref{eq:schroedinger:RS}) is the coordinate-invariant continuity equation $\p|\z_n^{(0)}|^2/\p t =-\f{1}{\sqrt{g_0}} \p_{\m}(\sqrt{g_0} J^{(0)\m})$, where the ionic current density
\begin{align}
J^{(0)\m} &= \mathrm{Re} \big[ \z_n^{(0)*} v^{\m} \z_n^{(0)} \big] \nn \\
&= |\z_n^{(0)}|^2 M_0^{\m\n} (\p_{\n} S - \hbar A_{\n}){.}
\end{align}
is defined in terms of the velocity operator 
\begin{align}
v^{\m} &= M_0^{\m\n} (p_{\n}- \hbar A_{\n}) {.}
\end{align}

Now we want to find the effective equation of motion for a classical particle with generalized coordinate $x$.  The kinetic energy term in the effective classical Hamiltonian derives from the second term of Eq.~(\ref{eq:schroedinger:S}) but can also be obtained as follows.  First, use the parametrization $\z_{n}^{(0)} = \sqrt{\rho} e^{i\hbar^{-1}S}$ to evaluate the expectation value of $\e T^{{\rm eff}(0)}$ as 
\begin{align}
\langle \e T^{{\rm eff}(0)} \rangle &= \int dx \sqrt{g_0} \f{\hbar^2}{2} \big[(\p_{\m}-i A_{\m}) \z_n^{(0)}\big]^* M_0^{\m\n} (\p_{\n}-i A_{\n}) \z_n^{(0)} \nn \\
&= \int dx \sqrt{g_0} |\z_n^{(0)}|^2 \f{1}{2} [ \p_{\m} S - \hbar A_{\m} ] M_0^{\m\n} [ \p_{\n} S - \hbar A_{\n} ] {.}
\label{eq:Tkin0:polar}
\end{align}
Defining the classical momentum $p_{\m}=\p_{\m} S$, we can read off the zeroth order classical ionic kinetic energy 
\begin{align}
\mathcal{T}^{{\rm ion}(0)}(x,p) &= \f{1}{2} M_0^{\m\n} (p_{\m} - \hbar A_{\m}) (p_{\n} - \hbar A_{\n}) {.}
\label{eq:Tion0}
\end{align}
The gauge potential does not play a role in the effective classical ionic equations of motion except if the original microscopic Hamiltonian contains external magnetic fields \cite{ceresoli2007,culpitt2021} or spin-orbit interactions \cite{mead1980}. 

The zeroth order effective ionic Hamiltonian is 
\begin{align}
\mathcal{H}^{{\rm ion}(0)}(x,p) &= \mathcal{T}^{{\rm ion}(0)}(x,p) + E_n(x) {.} 
\end{align}
After defining the velocity (a dot means differentiation with respect to time)
\begin{align}
\dot{x}^{\m} &= \f{\p \mathcal{H}^{{\rm ion}(0)}}{\p p_{\m}} \nn \\
&= M_0^{\m\n} (p_{\n} - \hbar A_{\n}) {,}
\end{align}
the zeroth order effective classical Lagrangian is found to be
\begin{align}
\mathcal{L}^{{\rm ion}(0)}(x,\dot{x}) &= p_{\m} \dot{x}^{\m} - \mathcal{H}^{{\rm ion}(0)}\nn \\
&= \f{1}{2} M_{0\m\n} \dot{x}^{\m} \dot{x}^{\n} - E_n(x) +\hbar A_{\m}\dot{x}^{\m} {.} 
\end{align}
The equation of motion that follows from the effective action $\mathcal{S}^{{\rm ion}(0)}[x] = \int dt \mathcal{L}^{{\rm ion}(0)}(x,\dot{x})$ is 
\begin{align}
M_{0\m\n} a^{\n} = \mathcal{F}_{\m}^{(0)} {,}
\end{align}
where $a^{\n} = \ddot{x}^{\n} + \Gamma_{0\r\s}^{\n} \dot{x}^{\r} \dot{x}^{\s}$ is the covariant acceleration, $\Gamma_{0\r\s}^{\n}$ was defined below Eq.~(\ref{eq:nablasq}), and $\mathcal{F}^{(0)}_{\m} = -\p_{\m} E_n$ is the zeroth order force.  The effective action $\mathcal{S}^{{\rm ion}(0)}[x]$ is equivalent to the effective action that was derived from the mixed quantum-classical action
\begin{align}
S[x,|\psi_{\rm elec}\rangle] &= \int dt \Big[ \f{1}{2} M_{0\m\n} \dot{x}^{\m} \dot{x}^{\n} + \langle \psi_{\rm elec}| i\hbar\f{d}{dt} - H(x)|\psi_{\rm elec}\rangle \Big]
\label{eq:action}
\end{align}
in the limit of low nuclear velocity [cf.~Eq.~(4.7) of Ref.~\onlinecite{chatzistavrakidis2026}, noting that $\e$ there corresponds to $\hbar$ here].  Equation~(\ref{eq:action}) is the action of an electron-ion system in which electrons are described by a time-dependent quantum state $|\psi_{\rm elec}\rangle$ and ions are described by a classical trajectory $x=x(t)$.

As we are interested in energy transfer, we now return to the quantum mechanical setting and calculate the rate of change of the ionic kinetic energy, which in the zeroth order adiabatic representation is expressed as
\begin{align}
\f{d\langle \e T\rangle}{dt} &= i \big< \psi^{(0)} \big|[ \mathcal{H}, \e T]^{(0)} \big|\psi^{(0)} \big> + \bigg< \f{\partial \e T}{\partial t} \bigg> {.}
\end{align}
The last term vanishes since the operator $T$ has no explicit time dependence.  Within the zeroth order adiabatic approximation, the formula evaluates to
\begin{align}
\f{d\langle \e T\rangle}{dt}\bigg|^{(0)} &= \int dx \sqrt{g_0} |\z_n^{(0)}|^2 F^{(0)}_{\m} V^{(0)\m} {,}
\label{eq:dT:0th order}
\end{align}
where $F^{(0)}_{\m} = -\p_{\m} (E_n + \e V_{geo})$ is the zeroth order force and $V^{(0)\m}=J^{(0)\m}/|\z_n^{(0)}|^2$ is the zeroth order velocity field.
Within the zeroth order adiabatic approximation, the wave function $\z_n^{(0)}$ is governed by an effective Schr\"odinger equation [Eq.~(\ref{eq: 0th order TDSE})].  Hence its energy is conserved.  If we interpret the kinetic energy as the ionic component of the energy and the potential energy as the electronic component, then Eq.~(\ref{eq:dT:0th order}) is a simple and intuitive formula for the rate of energy transfer between electrons and ions.  An interpretation along these lines, albeit with different quantities, was discussed in Ref.~\onlinecite{li2022}.

Now we look at energy transfer when the ions are treated classically.  Evaluating the equation of motion of $\mathcal{T}^{{\rm ion}(0)}$ in Eq.~(\ref{eq:Tion0}) gives 
\begin{align}
\f{d\mathcal{T}^{{\rm ion}(0)}}{dt} = \mathcal{F}^{(0)}_{\mu} \dot{x}^{\m} {,}
\label{eq:dTion:0}
\end{align}
This classical formula parallels the quantum formula in Eq.~(\ref{eq:dT:0th order}).  The classical formula can be obtained from the quantum formula under the assumption that $\zeta^{(0)}_n$ is a localized wave packet whose center moves along a trajectory $x(t)$, hence picking out $\dot{x}^{\m}(t)$ from the velocity field $V^{(0)\m}(x,t)$. The physical picture is of a classical particle moving in a curved space with metric $M_{0\m\n}$ and acted on by the gradient of the potential $E_n$.  

The leading order nonadiabatic corrections to the above results can be found by working within the first order adiabatic approximation.  However, there is a subtlety in defining the first order ionic kinetic energy.  
Because $G_1$ contains differential operators, the first order ionic kinetic energy is not simply the expectation value 
\begin{align}
\langle \e T \rangle|^{(1)} &= \int dx \sqrt{g_0} \z_n^{(1)*} \e T_{nn}^{(1)} \z_n^{(1)}
\label{eq:T1:expval}
\end{align}
with the operator
\begin{align}
\e T^{(1)} = \e T^{(0)} -i \e^2 [G_1,T]^{(0)} + \mathcal{O}(\e^3) {.} 
\label{eq:T(1):expansion}
\end{align} 
There are additional order $\e^2$ contributions from the other term, $H$, of the Hamiltonian, despite the fact that $H$ itself has nothing to do with ionic kinetic energy.  Specifically, in the expansion of $H$ in the first order adiabatic representation
\begin{align}
H^{(1)} &= H^{(0)} -i \e [G_1,H]^{(0)} - \f{1}{2} \e^2[G_1,[G_1,H]]^{(0)} + \mathcal{O}(\e^3) {,} 
\label{eq:H(1):expansion}
\end{align}
the operator in the third term should be considered as part of the first order effective ionic kinetic energy operator because its diagonal elements contain differential operators that act on $\z_n$.  Indeed, since $G_1$ was chosen such that $[G_1,H]_{ln}=-i T_{ln}$ for $l\neq n$, we immediately see that the complete order $\e^2$ nonadiabatic correction in $\mathcal{H}=\e T+H$ is the sum of the contributions from Eq.~(\ref{eq:T(1):expansion}) and Eq.~(\ref{eq:H(1):expansion}), i.e.
\begin{align}
-i \e^2 [G_1,T]^{(0)} - \f{1}{2} \e^2[G_1,[G_1,H]]^{(0)} &= -\f{i}{2} \e^2 [G_1,T]^{(0)} {.}
\label{eq:H:1st order correction}
\end{align}
We will see shortly that the operator on the right-hand side partitions, similarly to Eq.~(\ref{eq:Tnn:partitioning}), into kinetic energy and potential energy terms.

We now derive the effective time-dependent Schr\"odinger equation in the first order adiabatic representation.  In Sec.~\ref{sec:APT}, the time-independent Schr\"odinger equation was transformed to the first order adiabatic representation by choosing $G_1$ to make the off-diagonal Hamiltonian matrix elements vanish to order $\e$.  The first order adiabatic approximation consisted in neglecting the remaining off-diagonal coupling, of order $\e^2$, which gives a set of decoupled effective Schr\"odinger equations for the ionic wave functions.  The same procedure is successful in decoupling the components of the time-dependent Schr\"odinger equation in Eq.~(\ref{eq:adiabatic:0:td:A}) because its Hamiltonian is time independent.  In the first order adiabatic representation, the state $|\psi^{(1)}(x,t)\rangle$ maps to $[\zeta^{(1)}(x,t)] = (\ldots, 0,\z_{n}^{(1)}(x,t),0,\ldots)^T$ and the ionic wave function $\zeta_{n}^{(1)}$ satisfies the effective time-dependent Schr\"odinger equation
\begin{align}
i \p_{t} \zeta_{n}^{(1)} = \e T^{{\rm eff}(1)} \zeta_{n}^{(1)} + (E_n+\epsilon V_{geo}+\epsilon^2 V_{2,geo}) \zeta_{n}^{(1)} {.}
\label{eq: 1st order TDSE}
\end{align}
Nonadiabatic corrections make two changes with respect to the zeroth order Schr\"odinger equation in Eq.~(\ref{eq: 0th order TDSE}): they add the potential $\e^2 V_{2,geo}$ and they change $T^{{\rm eff}(0)}$ to $T^{{\rm eff}(1)}$, the {\it first order effective ionic kinetic energy operator}
\begin{align}
T^{{\rm eff}(1)} &= -\f{1}{2} \f{1}{\sqrt{g_0}} \big( \p_{\m} - i A^{(1)*}_{\m} \big) \sqrt{g_0} h^{(1)\m\n} \big( \p_{\n} - i A^{(1)}_{\n} \big) {.}
\label{eq:Tkin(1)}
\end{align}
These changes are due to the order $\e^2$ term in Eq.~(\ref{eq:H:1st order correction}), confirming that it partitions into kinetic energy and potential energy terms.

To shed light on the physical significance of the nonadiabatic corrections to $T^{{\rm eff}(1)}$, we again restore $\hbar$ and consider the classical limit $\hbar\rightarrow 0$.  Substituting the parametrization $\z_n^{(1)}=\sqrt{\rho}e^{i\hbar^{-1}S}$ into Eq.~(\ref{eq: 1st order TDSE}) and dividing by $\zeta_{n}^{(1)}$ yields
\begin{align}
\p_t S +&\f{1}{2} M^{(1)\m\n} (\p_{\m} S-\hbar A_{\m}^{(1)*}) (\p_{\n} S-\hbar A_{\n}^{(1)}) + E_n + \hbar^2 V_{geo,sc} + \hbar^4 V_{2,geo,sc} \nn\\
&= i\hbar \f{1}{\sqrt{\rho}} \p_t \sqrt{\rho} + i \f{\hbar}{2} \p_{\m} \ln \sqrt{\rho} M^{(1)\m\n} (\p_{\n} S-\hbar A_{\n}^{(1)}) + i \f{\hbar}{2} \f{1}{\sqrt{\rho}} (\p_{\m} S-\hbar A_{\m}^{(1)*}) M^{(1)\m\n} \p_{\n} \sqrt{\rho} \nn \\
&\quad +i \f{\hbar}{2} \f{1}{\sqrt{g_0}} \p_{\m} \big[ \sqrt{g_0} M^{(1)\m\n} (\p_{\n} S-\hbar A_{\n}^{(1)}) \big] + \f{\hbar^2}{2} \f{1}{\sqrt{\rho}} \f{1}{\sqrt{g_0}} \p_{\m} \big[ \sqrt{g_0} M^{(1)\m\n} \p_{\n} \sqrt{\rho} \big] {,}
\label{eq:schroedinger:1:RS}
\end{align}
where we have absorbed factors of $\e$ by defining the Hermitian mass tensor $M^{(1)}_{\m\n} = \e^{-1} h^{(1)}_{\m\n}$ and the second order geometric potential $V_{2,geo,sc}$ (in analogy to $V_{geo,sc}$). The real part gives
\begin{align}
\p_t S &+\f{1}{2} M^{(1)\m\n} (\p_{\m} S-\hbar A_{\m}^{(1)*}) (\p_{\n} S-\hbar A_{\n}^{(1)}) + E_n + \hbar^2 V_{geo,sc} + \hbar^4 V_{2,geo,sc}\nn\\
&-\f{\hbar^2}{2} \f{1}{\sqrt{\rho}} \f{1}{\sqrt{g_0}} \p_{\m} \big[ \sqrt{g_0}\, \mathrm{Re} M^{(1)\m\n} \p_{\n} \sqrt{\rho} \big] + \f{\hbar}{2} \f{1}{\sqrt{g_0}} \p_{\m} \big[ \sqrt{g_0}\, \mathrm{Im} [M^{(1)\m\n} (\p_{\n} S-\hbar A_{\n}^{(1)})] \big] =0  {.}
\end{align}
There are several changes with respect to the zeroth order result in Eq.~(\ref{eq:schroedinger:S}).  The Hermitian mass tensor $M^{(1)}_{\m\n}$ replaces $M_{0\m\n}$, and the complex-valued gauge potential $A_{\m}^{(1)}$ replaces $A_{\m}$.  The fifth term is like the so-called quantum potential \cite{bornemann1996,tully1998,agostini2018} but with $\mathrm{Re}M^{(1)\m\n}$ replacing $M_0^{\m\n}$.  The last term is of order $\hbar^3$ since $\mathrm{Im}M^{(1)\m\n}=\mathcal{O}(\hbar^2)$ and $\mathrm{Im}A_{\n}^{(1)}=\mathcal{O}(\hbar^2)$.  The dependence on the imaginary part of $A_{1,\m}$ enters at order $\hbar^4$.  The imaginary part of Eq.~(\ref{eq:schroedinger:1:RS}) is the continuity equation $\p|\z_n^{(1)}|^2/\p t = -\f{1}{\sqrt{g_0}} \p_{\m} [\sqrt{g_0} J^{(1)\m}]$ in the first order adiabatic representation with the ionic current density
\begin{align}
J^{(1)\m} &= \mathrm{Re} \big[ \z_n^{(1)*} M^{(1)\m\n} (\p_{\n}-\hbar A_{\n}^{(1)}) \z_n^{(1)} \big] {.}
\label{eq:J:1}
\end{align}

The expectation value of $T^{{\rm eff}(1)}$ can be written as
\begin{align}
\langle \e T^{{\rm eff}(1)} \rangle &= \int dx \sqrt{g_0} \f{\hbar^2}{2} \big[(\p_{\m}-i A_{\m}^{(1)}) \z_n^{(1)}\big]^* M^{(1)\m\n} (\p_{\n}-i A_{\n}^{(1)}) \z_n^{(1)} \nn \\
&= \int dx \sqrt{g_0} |\z_n^{(1)}|^2 \,\f{1}{2} [ \p_{\m} S - \hbar A_{\m}^{(1)*} +i\hbar \p_{\m} \ln \sqrt{\rho} ] M^{(1)\m\n} [ \p_{\n} S - \hbar A_{\n}^{(1)} -i\hbar \p_{\n} \ln \sqrt{\rho} ] {,}
\end{align}
from which we can read off the kinetic energy 
\begin{align}
\mathcal{T}^{{\rm ion}(1)}(x,p) &= \f{1}{2} (p_{\m} - \hbar A_{\m}^{(1)*}) M^{(1)\m\n} (p_{\n} - \hbar A_{\n}^{(1)}) \nn \\
&= \f{1}{2} \mathrm{Re} M^{(1)\m\n} (p_{\m} - \hbar \mathrm{Re} A_{\m}^{(1)}) (p_{\n} - \hbar \mathrm{Re} A_{\n}^{(1)}) \nn \\
&\quad + \hbar^3 (\p_{\m} \ln \sqrt{\rho}) M_0^{\m\r} \mathrm{Im} M_{1\r\s} M_0^{\s\n} (p_{\n} - \hbar \mathrm{Re} A_{\n})  {,}
\label{eq:Tion:1}
\end{align}
where $M_{1\m\n}=\e^{-1} h_{1\m\n}$.  It is interesting that the classical kinetic energy $\mathcal{T}^{{\rm ion}(1)}$ is built out of complex quantities, $M^{(1)\m\n}$ and $A_{\n}^{(1)}$, as seen in the first line of Eq.~(\ref{eq:Tion:1}).  In the second line, $\mathcal{T}^{{\rm ion}(1)}$ is broken down into pieces coming from the real and imaginary parts of $M^{(1)\m\n}$.  The piece containing the real part of $M^{(1)}_{\m\n}$ gives the kinetic energy to order $\hbar^2$ and has appeared in effective ionic \cite{littlejohn1993, goldhaber2005,scherrer2017,matyus2019,requist2023,littlejohn2024,requist2025,chatzistavrakidis2026} and electronic \cite{requist2010} Hamiltonians.  The second term is of order $\hbar^3$ and contains $\mathrm{Im} M_{1\m\n}$, which also appears in the first order effective kinetic energy operator [Eq.~(\ref{eq:Tkin(1)})] but has not previously been seen to play a role in the effective dynamics of classical ions.  The antisymmetric tensor $\mathrm{Im}M_{1\m\n}$ couples the zeroth order velocity vector $M_0^{\m\n}(p_{\n} - \hbar \mathrm{Re} A_{\n})$ to the vector $M_0^{\m\n} \p_{\n} \ln \sqrt{\rho}$, which depends on the ionic wave function through its density $\rho$.  Up until this point, it has been possible to express the classical ionic kinetic energy in the standard form $\f{1}{2}M(p-A)^2$ by absorbing the nonadiabatic effects into perturbed gauge potentials and mass tensors that depend on the adiabatic electronic state but not the ionic wave function. In particular, the ionic kinetic energy to order $\hbar^2$, the first term in Eq.~(\ref{eq:Tion:1}), has the standard $\f{1}{2}M(p-A)^2$ form but with a nonadiabatic gauge potential $\mathrm{Re} A_{\n}^{(1)}$ and a nonadiabatic mass tensor $\mathrm{Re}M^{(1)}_{\m\n}$.  Equation (\ref{eq:Tion:1}) shows that this structure cannot be preserved to order $\hbar^3$, since the second term depends on the ionic wave function.  Nevertheless, we will see that it is possible to express the ionic kinetic energy in the standard form $\f{1}{2}M\dot{x}^2$ to order $\hbar^3$. 

The first order effective classical Hamiltonian is 
\begin{align}
\mathcal{H}^{{\rm ion}(1)}(x,p) &= \mathcal{T}^{{\rm ion}(1)}(x,p) + E_n(x) +\hbar^2 V_{geo,sc}(x) - \f{\hbar^2}{2} \f{1}{\sqrt{\rho}} \mathrm{Re} M^{(1)\m\n} \nabla_{\m}^{(1)} \nabla_{\n}^{(1)} \sqrt{\rho} {.}
\label{eq:Hion:1}
\end{align}
The last term is the quantum potential term expressed in a coordinate-invariant form using the covariant derivative $\nabla^{(1)}$ compatible with $\mathrm{Re}M^{(1)}_{\m\n}$, i.e.
\begin{align}
\mathrm{Re} M^{(1)\m\n} \nabla_{\m}^{(1)} \nabla_{\n}^{(1)} \sqrt{\rho} &= \mathrm{Re} M^{(1)\m\n} \p_{\m} \p_{\n} \sqrt{\rho} - \mathrm{Re} M^{(1)\m\n} \Gamma_{\m\n}^{(1)\l} \p_{\l} \sqrt{\rho} {,} 
\end{align}
where $\Gamma^{(1)\l}_{\m\n}$ are the components of the Levi-Civita connection.  
In Eq.~(\ref{eq:Hion:1}), we have neglected the fourth order terms $\hbar^4 V_{2,geo,sc}$ and
\begin{align}
\f{\hbar^4}{4} \big[ (\p_{\m} \mathrm{Re} M_{1\r\s}) M_0^{\r\s} - (\p_{\m} M_{\r\s}) M_0^{\r\a} \mathrm{Re} M_{1\a\b} M_0^{\b\s} \big] \mathrm{Re} M^{(1)\m\n} \p_{\n} \ln \sqrt{\rho} {.}
\end{align}

The velocity, defined by 
\begin{align}
\dot{x}^{\l} &= \f{\p \mathcal{H}^{{\rm ion}(1)}}{\p p_{\l}} \nn \\
&= \mathrm{Re} M^{(1)\l\n} (p_{\n} - \hbar \mathrm{Re} A_{\n}^{(1)}) - \hbar^3 M_0^{\l\r} \mathrm{Im} M_{1\r\s} M_0^{\s\n} \p_{\n} \ln \sqrt{\rho} {,}
\label{eq:v}
\end{align}
is perturbed by an order $\hbar^3$ term depending on the antisymmetric tensor $M_0^{\m\r}\mathrm{Im}M_{1\r\s}M_0^{\s\n}$ and the logarithmic derivative of $\sqrt{\rho}$.  This correction will perturb the trajectory $x=x(t)$. With this definition of velocity, the ionic kinetic energy in Eq.~(\ref{eq:Tion:1}) takes on the standard form
\begin{align}
\mathcal{T}^{{\rm ion}(1)}(x,\dot{x}) &= \f{1}{2} M_{\m\n}^{(1)} \dot{x}^{\m} \dot{x}^{\n} {.}
\end{align}
The first order effective classical Lagrangian is
\begin{align}
\mathcal{L}^{{\rm ion}(1)}(x,\dot{x}) &= p_{\m} \dot{x}^{\m} - \mathcal{H}^{{\rm ion}(1)}\nn \\
&= \f{1}{2} M_{\m\n}^{(1)} \dot{x}^{\m} \dot{x}^{\n} - E_n(x) +\hbar A_{\m}\dot{x}^{\m} -\hbar^2 V_{geo,sc}(x) + \f{\hbar^2}{2} \f{1}{\sqrt{\rho}} \mathrm{Re}M^{(1)\m\n} \nabla_{\m} \nabla_{\n} \sqrt{\rho} {.}
\label{eq:L2}
\end{align}
The effective action $\mathcal{S}^{{\rm ion}(1)}[x] = \int \mathcal{L}^{{\rm ion}(1)}(x,\dot{x})dt$ is the same as the effective action derived from the mixed quantum-classical action in Eq.~(\ref{eq:action}) [cf.~Eq.~(4.7) in Ref.~\onlinecite{chatzistavrakidis2026}, recalling that $\e$ there corresponds to $\hbar$ here] in the limit of slow ionic velocity, except for the presence of the potentials $\hbar^2 V_{geo}$ and $-\f{\hbar^2}{2} \f{1}{\sqrt{\rho}} \mathrm{Re}M^{(1)\m\n} \nabla_{\m} \nabla_{\n} \sqrt{\rho}$.  The Euler-Lagrange equation for $\mathcal{S}^{{\rm ion}(1)}$ is not a closed equation of motion for the trajectory of the particle because it contains the $\rho$-dependent quantum potential.  $\rho$ is in turn determined by the continuity equation.

Returning to the quantum mechanical setting, we note that since $\e T^{{\rm eff}(1)}$ is the kinetic energy operator of a self-contained Schr\"odinger equation [Eq.~(\ref{eq: 1st order TDSE})], we can calculate its rate of change directly from the Heisenberg equation of motion with the Hamiltonian $\mathcal{H}^{{\rm eff}(1)}$ that appears in that Schr\"odinger equation rather than the full electron-ion Hamiltonian $\mathcal{H}^{(1)}$. The result is
\begin{align}
\f{d\langle \e T^{{\rm eff}(1)}\rangle}{dt} &= i \big< \big[ \mathcal{H}^{{\rm eff}(1)}_{nn}, \e T^{{\rm eff}(1)} \big] \big> \nn \\
&= \int dx \sqrt{g_0} |\z_n^{(1)}|^2 F_{\m}^{(1)} V^{(1)\m} {,}
\label{eq:dTeff:1}
\end{align}
where $V^{(1)\m}=J^{(1)\m}/|\z_n^{(1)}|^2$ and $F^{(1)}_{\m} = -\p_{\m} (E_n + \e V_{geo,n} + \e^2 V_{2,geo})$ is the first order force.  As the ionic wave function evolves on the adiabatic potential energy surface, there is an exchange of energy between kinetic $\langle \e T^{{\rm eff}(1)}\rangle$ and potential $\int dx \sqrt{g_0}|\z_n^{(1)}|^2 (E_n + \e V_{geo,n} + \e^2 V_{2,geo})$ energy.  The above results quantify the nonadiabatic corrections to energy transfer when the ions are treated quantum mechanically and the electrons remain in a single adiabatic state. 

As an aside, we show that the $G_1$ in Eq.~(\ref{eq:G1}) reduces to the $G_1$ in Eq.~(3.28) of Ref.~\onlinecite{chatzistavrakidis2026}. Consider the quantity
\begin{align}
\f{G_{1,ln} \z_n}{\z_n} &= \f{i}{E_l-E_n} \f{T_{ln} \z_n}{\z_n} -i g_0^{\r\s} \f{\p_{\r}E_n}{(E_l-E_n)^2} (-iA_{\s,ln}) {.}
\end{align}
Keeping only the leading order terms, i.e.~those containing derivatives acting on $\z_n$, gives
\begin{align}
\f{G_{1,ln} \z_n}{\z_n} &= \f{\langle l|\p_{\m} n\rangle}{E_l-E_n} \f{v^{\m} \z_n}{\z_n} {,} 
\end{align}
which reduces to the $G_1$ in Eq.~(3.28) of Ref.~\onlinecite{chatzistavrakidis2026} after replacing $v^{\m}\z_n/\z_n$ by $\dot{x}^{\m}$.

\subsection{Effective Hamiltonian under slow external driving \label{ssec:slow-driving}}

We now turn our attention to the second dynamical problem and consider an electron-ion system that is driven by slowly changing external parameters. The system obeys the time-dependent Schr\"odinger equation 
\begin{align}
i \f{d}{dt} |\psi(x,t)\rangle = \mathcal{H}(x,\l) |\psi(x,t)\rangle 
\label{eq:td:schroedinger:td:H}
\end{align}
with a Hamiltonian that depends on external parameters $\l^{a}=\l^{a}(\e_0 t)$.  The adiabatic parameter $\epsilon_0$, which is distinct from the adiabatic parameter $\e$, controls the rate of change of the parameters.  In general, both the ionic kinetic energy $\e T$ and the Hamiltonian $H$ may depend on $\l$, as is the case for an external magnetic field.  We will assume here that the ionic kinetic energy is $\l$-independent.  This assumption can be relaxed.  We seek an approximate solution to the time-dependent Schr\"odinger equation in the limit that $\epsilon_0$ tends to zero.  By changing the time variable to $t'=\epsilon_0 t$ and defining $\lambda'=\lambda'(t')=\lambda(\e_0 t)$, we obtain 
\begin{align}
    i\epsilon_0\frac{d}{dt'}|\psi(x,t')\rangle = \mathcal{H}(x,\l')|\psi(x,t')\rangle.
    \label{eq:td:schroedinger:scaled}
\end{align}
This is the standard form of the Schr\"odinger equation in the time-dependent version of adiabatic perturbation theory.  Equation (\ref{eq:td:schroedinger:scaled}) is a singularly perturbed differential equation in the $\epsilon_0\rightarrow 0$ limit.  From now on, we will drop the primes.

We will approximate the dynamics of the electron-ion system when $\e_0$ and $\e$ are small.  
There are several dynamical regimes that can be considered, depending on the relative orders of magnitude of $\e_0$ and $\e$.  The eigenvalue equation
\begin{align}
    H(x,\l) |n(x,\l)\rangle = E_n(x,\l) |n(x,\l)\rangle
    \label{eq:BO:eigenvalue:td}
\end{align}
defines adiabatic electronic eigenstates with parametric dependence on the external parameters $\l^{a}$ in addition to the ionic coordinates $x^{\m}$.  An electron-ion state can be expanded in this adiabatic representation as
\begin{align}
    |\psi(x,t)\rangle = \sum_n \zeta_n(x,t)|n(x,\l)\rangle.
    \label{eq:psi:td}
\end{align}
Substituting this expansion into the time-dependent Schr\"odinger equation 
and projecting onto $|l\rangle$ yields the coupled equations
\begin{align}    \sum_n i\e_0 D_{t,ln} \zeta_{n} = -\frac{\epsilon}{2} \sum_{mn} \f{1}{\sqrt{g_0}} (\partial_\mu \delta_{lm} -iA_{\mu,lm}) \sqrt{g_0} g^{\mu\nu}_0 (\partial_\nu \delta_{mn} -iA_{\nu,mn}) \zeta_{n} + E_l\zeta_{l},
\label{eq:schrodinger:coupled}
\end{align}
where $D_{t,ln}=\p_t \delta_{ln} - i A_{t,ln}$ with $A_{t,ln} = i\langle l|\partial_t n\rangle$.  We can write these equations as
\begin{align}
i\e_0 D_{t,ll} \z_l &= \sum_n \mathcal{H}_{ln}^{(0)}(x,\l) \z_n {,}
\label{eq:schrodinger:coupled:H}
\end{align}
in which the Hamiltonian matrix elements 
\begin{align}
\mathcal{H}_{ln}^{(0)}(x,\l) &= -\frac{\epsilon}{2} \sum_{m} \f{1}{\sqrt{g_0}} (\partial_\mu -iA_{\mu,lm}) \sqrt{g_0} g^{\mu\nu}_0 (\partial_\nu \delta_{mn} -iA_{\nu,mn}) + E_l(x,\l) \delta_{ln} - \e_0 A_{t,ln}
\label{eq:H:xlambda}
\end{align}
have the same form as those in the static case [cf.~Eq.~(\ref{eq:H0})], except $E_n$ and $A_{\m,ln}$ have become $\l$-dependent and there is an additional term $-\e_0 A_{t,ln}$.  There are now two distinct sources of nonadiabatic coupling:  the nonadiabatic coupling $\e T_{ln}$ induced by the ionic kinetic energy operator, and the non\-adiabatic coupling $\e_0 A_{t,ln}$ induced by the time derivative.  If we neglect both types of nonadiabatic coupling, we obtain the set of decoupled effective Schr\"odinger equations 
\begin{align}
    i \e_0 D_{t,nn} \zeta_{n}^{(0)} = \mathcal{H}_{nn}^{{\rm eff}(0)}(x,\l) \zeta_{n}^{(0)} {,}
\label{eq:schrodinger:zeta:td}
\end{align}
with the effective Hamiltonian
\begin{align}
\mathcal{H}_{nn}^{{\rm eff}(0)}(x,\l) = \e T^{{\rm eff}(0)}(\l) + E_n(x,\l) + \e V_{geo}(x,\l) {.}
\end{align}
The ionic kinetic energy operator $\e T^{{\rm eff}(0)}(\l)$ is potentially $\lambda$-dependent.

By separating electronic and ionic variables, the adiabatic approximation has reduced the complexity of the joint electron-ion Schr\"odinger equation in Eq.~(\ref{eq:td:schroedinger:scaled}) to the much lower complexity of the set of decoupled effective ionic Schr\"odinger equations in Eq.~(\ref{eq:schrodinger:zeta:td}).  Except in the simplest cases, the dimension of the nuclear configuration space is still too large for the effective ionic Schr\"odinger equations to be solvable without further approximations.  Once we obtain an approximate or numerical solution to Eq.~(\ref{eq:schrodinger:zeta:td}), we can evaluate the time evolution of any electronic or ionic observable in the zeroth order adiabatic approximation.

Our objective now is to account for nonadiabatic effects without significantly increasing the computational difficulty of the resulting effective ionic equations.  Our strategy will be to apply a sequence of unitary transformations, but they will be different than those in Sec.~\ref{sec:APT}.  Applying first a time-dependent unitary transformation of the form
\begin{align}
V_1 &= e^{i\e_0 g_1} {}
\end{align}
transforms the Hamiltonian to
\begin{align}
\mathcal{H}^{[1](0)} &= V_1^{\dag} \mathcal{H}^{(0)} V_1 - i \e V_1^{\dag} \dot{V}_1 {.}
\end{align}
The notation has been supplemented: an integer in square brackets denotes the order of the representation with respect to transformations involving $\e_0$, and an integer in parentheses denotes the order of the representation with respect to transformations involving $\e$.  Expanding in $\e_0$ yields
\begin{align}
\mathcal{H}^{[1](0)} &= H - i \e_0 [g_1,H] - \f{1}{2}\e_0^2 [g_1,[g_1,H]] + \f{i}{6} \e_0^3 [g_1,[g_1,[g_1,H]]] + \f{1}{24} \e_0^4 [g_1,[g_1,[g_1,[g_1,H]]]] \nn \\
&\quad - \e_0 A_t + i \e_0^2 [g_1,A_t] + \f{1}{2} \e_0^3 [g_1,[g_1,A_t]] - \f{i}{6} \e_0^4 [g_1,[g_1,[g_1,A_t]]] \nn \\
&\quad +\e T    -i \e_0 \e [g_1,T]  -\f{1}{2} \e_0^2 \e [g_1,[g_1,T]] + \e_0^2 \dot{g}_1 - \f{i}{2} \e_0^3 [g_1,\dot{g}_1] - \f{1}{6} \e_0^4 [g_1,[g_1,\dot{g}_1]]
\end{align}
up to higher order terms.  The condition that $g_1$ must satisfy to eliminate the order $\e_0$ off-diagonal elements is 
\begin{align}
-i [g_1,H]_{ln} - A_{t,ln} &= 0 {}
\end{align}
for $l\neq n$.  The solution is
\begin{align}
g_{1,ln} &= \f{F_{t,ln}}{E_{ln}}
\end{align}
with $F_{t,ln} = -i A_{t,ln}$. We set the diagonal elements of $g_1$ to zero, which is a gauge choice.
With this choice of $g_1$, the Hamiltonian simplifies to
\begin{align}
\mathcal{H}^{[1](0)} &= H + \f{i}{2}\e_0^2 [g_1,A_{t,od}] + \f{1}{3} \e_0^3 [g_1,[g_1,A_{t,od}]] - \f{i}{8} \e_0^4 [g_1,[g_1,[g_1,A_{t,od}]]] \nn \\
&\quad - \e_0 A_{t,d} + i \e_0^2 [g_1,A_{t,d}] + \f{1}{2} \e_0^3 [g_1,[g_1,A_{t,d}]] - \f{i}{6} \e_0^4 [g_1,[g_1,[g_1,A_{t,d}]]] \nn \\
&\quad +\e T    -i \e_0 \e [g_1,T]  -\f{1}{2} \e_0^2 \e [g_1,[g_1,T]] \nn \\
&\quad + \e_0^2 \dot{g}_1 - \f{i}{2} \e_0^3 [g_1,\dot{g}_1] - \f{1}{6} \e_0^4 [g_1,[g_1,\dot{g}_1]] {.}
\end{align}
Here the subscript $od$ indicates the off-diagonal part of the operator and $d$ indicates the diagonal part. The second step is to apply a time-dependent unitary transformation of the form
\begin{align}
U_1 = e^{i\e G_1} {,}
\end{align}
which transforms the Hamiltonian to
\begin{align}
\mathcal{H}^{[1](1)} &= U_1^{\dag} \mathcal{H}^{[1](0)} U_1 -i \e_0 U_1^{\dag} \dot{U}_1 {.}
\end{align}
This produces many terms.  Our objective is to choose $G_1$ such that the off-diagonal coupling in $\mathcal{H}^{[1](1)}$ becomes of higher order in $\e$.  To proceed we need to know the relative orders of magnitude of $\e_0$ and $\e$.  We consider the special case $\e=\e_0^2$.  This gives us an equation in which the small parameter $\e_0$ enters in the same way as the factors of $\hbar$ associated with $d/dt$ and $T$.  In view of this choice, we rename the Hamiltonian $\mathcal{H}^{[1](0)}$ as $\mathcal{H}^{(1)}$, the unitary transformation $U_1$ as $V_2$, and the generator $G_1$ as $g_2$.  With these identifications, the Hamiltonian $\mathcal{H}^{(1)}$ becomes
\begin{align}
\mathcal{H}^{(1)} &= H - \e_0 A_{t,d} + \f{i}{2}\e_0^2 [g_1,A_{t,od}] + i \e_0^2 [g_1,A_{t,d}] +\e_0^2 T + \e_0^2 \dot{g}_1 \nn \\
&\quad + \f{1}{3} \e_0^3 [g_1,[g_1,A_{t,od}]] + \f{1}{2} \e_0^3 [g_1,[g_1,A_{t,d}]] -i \e_0^3 [g_1,T] - \f{i}{2} \e_0^3 [g_1,\dot{g}_1] \nn \\
&\quad - \f{i}{8} \e_0^4 [g_1,[g_1,[g_1,A_{t,od}]]] - \f{i}{6} \e_0^4 [g_1,[g_1,[g_1,A_{t,d}]]] -\f{1}{2} \e_0^4 [g_1,[g_1,T]] - \f{1}{6} \e_0^4 [g_1,[g_1,\dot{g}_1]] {.}
\end{align}
Applying the transformation $V_2$ gives the Hamiltonian 
\begin{align}
\mathcal{H}^{(2)} 
&= H - \e_0 A_{t,d} -i \e_0^2 [g_2,H] + \f{i}{2} \e_0^2 [g_1,A_{t,od}] + i \e_0^2 [g_1,A_{t,d}] + \e_0^2 T + \e_0^2 \dot{g}_1 \nn \\
&\quad +i \e_0^3 [g_2,A_{t,d}] + \f{1}{3} \e_0^3 [g_1,[g_1,A_{t,od}]] + \f{1}{2} \e_0^3 [g_1,[g_1,A_{t,d}]] -i \e_0^3 [g_1,T] - \f{i}{2} \e_0^3 [g_1,\dot{g}_1] \nn \\
&\quad -\f{1}{2} \e_0^4 [g_2,[g_2,H]] + \f{1}{2} \e_0^4 [g_2,[g_1,A_{t,od}]] + \e_0^4 [g_2,[g_1,A_{t,d}]] -i \e_0^4 [g_2,T] - i \e_0^4 [g_2,\dot{g}_1] \nn \\
&\quad - \f{i}{8} \e_0^4 [g_1,[g_1,[g_1,A_{t,od}]]] - \f{i}{6} \e_0^4 [g_1,[g_1,[g_1,A_{t,d}]]] -\f{1}{2} \e_0^4 [g_1,[g_1,T]] - \f{1}{6} \e_0^4 [g_1,[g_1,\dot{g}_1]] {.}
\end{align}
The condition that $g_2$ must satisfy to remove the order $\e_0^2$ off-diagonal terms is
\begin{align}
-i[g_2,H]_{ln} + T_{ln} +\f{i}{2} [g_1,A_{t,od}]_{ln} + i [g_1,A_{t,d}]_{ln} + \dot{g}_{1,ln} = 0 {}
\end{align}
for $l\neq n$.  
The solution 
\begin{align}
g_{2,ln} &= \f{i}{E_l-E_n} T_{ln} - i g_0^{\m\n} \f{\p_{\m} E_n}{(E_l-E_n)^2}(-iA_{\n,ln}) + G_{2,ln}
\end{align}
is similar to Eq.~(\ref{eq:G1}) except for the extra term $G_2$ originating from the time dependence of the $\l$ parameters.  The expression for $G_2$ is given in Eq.~(3.55) of Ref.~\onlinecite{chatzistavrakidis2026} with the parameters $x^{\m}$ replaced by $\l^{a}$.   The next step is to perform a further transformation with a unitary operator of the form
\begin{align}
V_3 = e^{i\e_0^3 g_3} {}
\end{align}
to remove the order $\e_0^3$ off-diagonal coupling.  However, we do not need to know $g_3$ in order to write down the effective Hamiltonian in the second order adiabatic representation.  After neglecting the order $\e_0^3$ off-diagonal coupling, a general diagonal element (to order $\e_0^3$) is
\begin{align}
\mathcal{H}^{{\rm eff}(2)}_{nn} 
&= E_n - \e_0 A_{t,nn} + \e_0^2 T_{nn} + \f{i}{2} \e_0^2 [g_1,A_{t,od}]_{nn}  \nn \\
&\quad + \f{1}{3} \e_0^3 [g_1,[g_1,A_{t,od}]]_{nn} + \f{1}{2} \e_0^3 [g_1,[g_1,A_{t,d}]]_{nn} - \f{i}{2} \e_0^3 [g_1,\dot{g}_1]_{nn} -i \e_0^3 [g_1,T]_{nn} {.}
\label{eq:Heff:2}
\end{align}
The first term and the third term are familiar from the zeroth order effective Hamiltonian [see Eq.~(\ref{eq:Tnn:partitioning})].  The order $\e_0$ and order $\e_0^2$ terms associated with the external driving fields are
\begin{align}
- \e_0 A_{t,nn} &= -\e_0 A_{a,nn} \dot{\lambda}^{a} \nn \\
\f{i}{2} \e_0^2 [g_1,A_{t,od}]_{nn} &= -\f{1}{2} M_{1ab} \dot{\l}^{a} \dot{\l}^{b} {,}
\end{align}
where
\begin{align}
M_{1ab} = -2 \mathrm{Re} \langle D_{a} n|(E_n-H)^{-1}|D_{b} n\rangle
\end{align}
is an induced mass tensor associated with the $\l$ parameters, $D_a = \p/\p \l^a +i A_{a,nn}$ is the gauge covariant derivative with respect to $\l^a$, and $A_{a,nn} = i\langle n|(\p/\p \l^a) n\rangle$ is the gauge potential. 
The following three order $\e_0^3$ terms can be written compactly as
\begin{align}
\f{1}{3} \e_0^3 [g_1,[g_1,A_{t,od}]]_{nn} + \f{1}{2} \e_0^3 [g_1,[g_1,A_{t,d}]]_{nn} - \f{i}{2} \e_0^3 [g_1,\dot{g}_1]_{nn} &= 
-\e_0^3 \Big[ \f{1}{2} \omega_{2ab} \dot{\l}^{a} \ddot{\l}^{b} + \f{1}{6} \gamma_{2abc} \dot{\l}^{a} \dot{\l}^{b} \dot{\l}^{c} \Big] 
\label{eq:third}
\end{align}
with
\begin{align}
\omega_{2\m\n} &= -2\mathrm{Im} \langle D_{\m} n|(E_n-H)^{-2}|D_{\n} n\rangle, \nn \\
\gamma_{2\l\m\n} &= -6\mathrm{Im} \langle D_{(\l} n|(E_n-H)^{-2} |D_{\m} D_{\n)} n\rangle \nn \\[4pt]
&\quad - 6\mathrm{Im} \langle D_{(\l} n|(E_n-H)^{-1} \partial_{\m} (E_n-H)^{-1} |D_{\n)} n\rangle {.}
\label{eq:omega:gamma}
\end{align}
The parentheses in the subscripts denote the symmetrization of the indices.  Therefore, $\gamma_{2\l\m\n}$ is a fully symmetric object.  The third order terms in Eq.~(\ref{eq:third}) and the geometric objects in Eq.~(\ref{eq:omega:gamma}) were found in the effective quantum Hamiltonian of a quantum-classical system in the low velocity limit [see Eqs.~(3.37) and (3.38) of Ref.~\onlinecite{chatzistavrakidis2026}].  The final third order term in Eq.~(\ref{eq:Heff:2}) is
\begin{align}
-i \e_0^3 [g_1,T]_{nn} {.}
\end{align}
This term describes how the ionic kinetic energy operator is altered by the external driving.  As in Sec.~\ref{ssec:slow-ions}, the effective Hamiltonian can be used to define a higher order effective ionic kinetic energy operator, now $\l$-dependent, and evaluate its time rate of change.

The energy transfer formulas we have derived in this section are valid in the adiabatic regime.  Many experiments probe the dynamical response to a perturbation that is weak but not slowly varying. We can quantify energy transfer in those cases using linear response theory, which is the topic of the next section.  Since adiabatic perturbation theory produces an effective Hamiltonian for the full electron-ion system, it can also be used to evaluate energy transfer in systems described by density matrices, e.g.~in finite temperature and nonequilibrium systems.

\section{Nonadiabatic corrections to linear response functions \label{sec:LR}}

The response of a system to a weak external field can be evaluated in linear response theory.  If the system is initially in a mixed state, such as a thermal ensemble, described by the density matrix $\rho_0 = \sum_n w_n |\psi_n\rangle \langle \psi_n|$, the expectation value of an operator $O$ is
\begin{align}
    \langle O\rangle_0 = \sum_n w_n\langle\psi_n|O|\psi_n\rangle {.}
\end{align}
To linear order, the change of the expectation value of the operator $O$ due to a perturbation $f(t)B$ caused by an external time-dependent field $f(t)$ that couples to the system via the operator $B$ is
\begin{align}
    \langle O\rangle_1(t) 
    = \int_{-\infty}^\infty \text{d}t'\,\chi_{OB}(t-t')f(t').
\end{align}
The causal response function $\chi_{OB}(t-t') = -i\theta(t-t')\langle [O_H(t),B_H(t') ] \rangle$ contains operators in the Heisenberg picture, i.e.~$O_H(t) = e^{iHt}Oe^{-iHt}$ (see e.g.~Ref.~\onlinecite{Giuliani_Vignale_2005}).  In the frequency domain, 
\begin{align}
    \langle O\rangle_1(\omega) = \chi_{OB}(\omega)f(\omega) {.}
\end{align}
The key quantity is the Fourier transform of the response function
\begin{align}
    \chi_{OB}(\omega) &= \int_{-\infty}^{\infty} d(t-t') \chi_{OB}(t-t') \nn \\
    &= \lim_{\eta\rightarrow0^+}\sum_{LN}\frac{w_N-w_L}{\omega-(\Lambda_L-\Lambda_N) + i\eta}O_{NL}B_{LN}.
    \label{eq:chiOB}
\end{align}
In electron-ion systems, the summation in Eq.~(\ref{eq:chiOB}) must be taken over a very large set of joint electron-ion states, which is computationally challenging.  However, the response of the ions is usually weaker than the response of the electrons.  Indeed, linear response functions involving electronic observables are often evaluated with the ions held fixed at positions corresponding to a reference point $x_0$, the {\it static ions} approximation,
\begin{align}
    \chi_{OB}^{\textrm{static ions}}(\omega) = \lim_{\eta\rightarrow0^+} \sum_{ln}\frac{w_{n}-w_{l}}{\omega-[E_{l}(x_0) - E_{n}(x_0)] + i\eta} O_{nl}(x_0) B_{ln}(x_0) {.}
    \label{eq:chiOB:static ions}
\end{align}
Now the summations only run over electronic states, but the response function is missing nonadiabatic corrections induced by the ionic motion.  We now show how these nonadiabatic corrections can be reintroduced to any order in $\e$ using adiabatic perturbation theory.

The linear response function can be evaluated in any representation.  To capture non\-adiabatic corrections to order $\e^p$, we evaluate the linear response function in the $p$th order adiabatic representation.  As explained in Sec.~\ref{sec:APT}, adiabatic perturbation theory seeks to decouple the electronic states.  Neglecting the order $\e^{p}$ off-diagonal coupling gives the $p$th order effective Hamiltonian $\mathcal{H}^{{\rm eff}(p)}$.  Its eigenstates $|\psi_{n\g}^{(p)}(x)\rangle$ and eigenenergies $\Lambda_{n\g}^{(p)}$, which are labeled by separate electronic and ionic quantum numbers, can be used to evaluate the linear response function in Eq.~(\ref{eq:chiOB}) as 
\begin{align}
    \chi_{OB}(\omega) = \lim_{\eta\rightarrow0^+}\sum_{\substack{l,\a\\n,\g}}\frac{w_{n\g}-w_{l\a}}{\omega-(\Lambda_{l\a}^{(p)} - \Lambda_{n\g}^{(p)}) + i\eta} \langle\zeta_{n\g}|O_{nl}|\zeta_{l\a}\rangle^{(p)} \langle\zeta_{l\a}|B_{ln}|\zeta_{n\g}\rangle^{(p)}.
    \label{eq:chiOB:p}
\end{align}
The bra-ket notation denotes the matrix elements
\begin{align}
\langle\zeta_{l\a}|B_{ln}|\zeta_{n\g}\rangle^{(p)} = \int dx \zeta_{l\a}^{(p)*}(x) B_{ln}^{(p)}(x) \zeta_{n\g}^{(p)}(x) {.}
\label{eq:matrix:elements:B}
\end{align}
Let us show how to evaluate the matrix elements in Eq.~(\ref{eq:chiOB:p}) in the first order adiabatic representation ($p=1$).  Using a Baker-Campbell-Hausdorff-type formula to expand $O^{(1)}$ and $B^{(1)}$ 
as in Eq.~(\ref{eq:H(1)}), we get a zeroth order term 
\begin{align}
\langle\zeta_{n\g}^{(1)}|O_{nl}|\zeta_{l\a}^{(1)}\rangle\langle\zeta_{l\a}^{(1)}|B_{ln}|\zeta_{n\g}^{(1)}\rangle {}
\end{align}
and two first order corrections
\begin{align}
-i\e \langle\zeta_{n\g}^{(1)}|[G_1,O]_{nl}|\zeta_{l\a}^{(1)}\rangle\langle\zeta_{l\a}^{(1)}|B_{ln}|\zeta_{n\g}^{(1)}\rangle -i\e \langle\zeta_{n\g}^{(1)}|O_{nl}|\zeta_{l\a}^{(1)}\rangle\langle\zeta_{l\a}^{(1)}|[G_1,B]_{ln}|\zeta_{n\g}^{(1)}\rangle {.}
\label{eq:first order correction}
\end{align}
Since we are working to first order in adiabatic perturbation theory, we keep only terms to first order in $\epsilon$.  Higher accuracy can be obtained by evaluating the response function in a higher order adiabatic representation.  The ionic wave functions $|\zeta_{n\g}^{(1)}\rangle$ are the eigenstates of the first order effective Hamiltonian $\mathcal{H}^{{\rm eff}(1)}_{nn}$.  They contain additional $\e$ dependence.

\section{Model system \label{sec:model}}

In order to test adiabatic perturbation theory for energy transfer and charge transport in a concrete system where numerical calculations can be performed to a high level of accuracy without uncontrolled approximations, we need to construct a sufficiently simple yet nontrivial model system.  We want it to be feasible to converge our numerical calculations with respect to the size of the basis set. The system should be representative of extended condensed matter systems with translational symmetry. We choose the system to be one dimensional in order to keep numerical computations manageable.  An important property of the Hamiltonian is that it contains differential kinetic energy operators for all particles. 

This kind of model system will allow us to test approximations based on the adiabatic small parameter $\epsilon$, without the confounding factor of additional approximations, such as mean-field-type approximations for the electronic structure.  We study external perturbations that couple to the system through a vector potential and do not break the translational symmetry of the model.

The model system we choose is composed of $N$ positively charged ions and $N_e$ electrons situated on a one dimensional manifold of length $L$ with periodic boundary conditions. Each ion represents a nucleus with charge $+Z_i e$ together with $Z_i-1$ core electrons, assumed to be inactive, so that each ion has an overall charge of $+e$. The coordinates of the ions will be denoted by $x_{ni}$ and the coordinates of the electrons by $r_{ei}$. We assume the ions are distinguishable particles. The system will be perturbed by applying a time-dependent magnetic flux generated by a gauge potential $A(t)$.

The Hamiltonian of our model electron-ion system is
\begin{equation}
\hat{\mathcal{H}}_{ring}(t) = \hat{T}_n(t) + \hat{T}_e(t) + \hat{V}_{nn} + \hat{V}_{ee} + \hat{V}_{en}
\end{equation}
with
\begin{align}
\hat{T}_n &= \sum_{i=1}^N \frac{(\hat{P}_i-e A(\hat{x}_{ni},t))^2}{2M_i} \nn \\
\hat{T}_e &= \sum_{i=1}^{N_e} \frac{(\hat{p}_i+e A(\hat{r}_{ei},t))^2}{2m} \nn \\
\hat{V}_{nn} &= \frac{1}{2} \sum_{i\neq j} V_{nn}(\hat{x}_{ni}-\hat{x}_{nj}) \nn \\
\hat{V}_{ee} &= \frac{1}{2} \sum_{i\neq j} V_{ee}(\hat{r}_{ei}-\hat{r}_{ej}) \nn \\
\hat{V}_{en} &= \sum_{i,j} V_{en}(\hat{r}_{ei}-\hat{x}_{nj}).
\end{align}
We can express the ionic kinetic energy as
\begin{align}
\hat{T}_n &= \sum_{i=1}^N \frac{1}{2M_i}\Big[-i\hbar \frac{d}{dx_{ni}}-e A(t)\Big]^2,
\end{align}
where we have assumed that the vector potential is independent of $x$, which can always be achieved in one-dimensional systems by a gauge transformation.
If we set $M_i=M$, the ionic kinetic energy can be further expressed as
\begin{align}
\hat{T}_n &= \frac{\hbar^2}{2M} \sum_{i=1}^N \bigg[-i\frac{d}{dx_{ni}}-\frac{2\pi A(t)}{\Phi_0}\bigg]^2 \nn \\
&= \frac{\hbar^2}{2M} \sum_{i=1}^N \Big[-i\frac{d}{dx_{ni}}-b(t)\Big]^2,
\end{align}
where $\Phi_0 = h/e$ is the Dirac flux quantum. 

We perform a change of variables to split off the center-of-mass motion.\cite{vanleeuwen2004, Sutcliffe_coordinates} It is done in two steps, first we make a change of variables to isolate the ionic center-of-mass coordinate, then we make a further change of variables to isolate the total center-of-mass coordinate.

The coordinate transformations of the first step are
\begin{align}
    x_{ncm} &= \frac{1}{M_{n,tot}} \sum_{i=1}^N M_i x_{ni} \\
    x_i &= \sum_j x_{nj} V_{ji}
\end{align}
with the electronic variables left unchanged. $V$ is an $N\times N$ matrix with
\begin{align}
V_{jN} &= \frac{M_j}{M_{n,tot}}
\end{align}
that satisfies
\begin{align}
\sum_{j=1}^N V_{ji} = 0 \;\; \mathrm{for} \;\; i<N.
\label{eq:V:condition}
\end{align}
The columns of $V$ are collective ionic coordinates of the ions. We can always choose them to be mutually orthogonal. For example, for $N=4$, we can choose
\begin{equation}
    V = \left[\begin{array}{cccc}
        -\frac{1}{2} & -\frac{1}{\sqrt{2}} & 0 & \frac{1}{4} \\
        \frac{1}{2} & 0 & -\frac{1}{\sqrt{2}} & \frac{1}{4} \\
        -\frac{1}{2} & \frac{1}{\sqrt{2}} & 0 & \frac{1}{4} \\
        \frac{1}{2} & 0 & \frac{1}{\sqrt{2}} & \frac{1}{4}
    \end{array} \right] {.}
\end{equation}
We have normalised the first $N-1$ column vectors to 1. We can consider $x_{ncm}$ as the $N$th $x_j$ variable, i.e.~$x_{ncm} = x_N$. With this choice of $V$, the new $x$ coordinates are the amplitudes of normal mode vibrations. They are the phonons of our system when it crystallizes into a periodic structure.

The coordinate transformations of the second step are
\begin{align}
r_i &= r_{ei} - x_{ncm} \label{eq: coordinate transform}\\
x_{cm} &= x_{ncm} + \frac{m}{M_{tot}} \sum_{i=1}^{N_e} r_i.
\end{align}
The inverse of the two-step coordinate transformation is
\begin{align}
r_{ei} &= r_{i} + x_{ncm}(\{r_i\},x_{cm}) \\ 
x_{ni} 
&= \sum_{j=1}^{N-1} x_j (V^{-1})_{ji} + x_{ncm}(\{r_i\},x_{cm})  {,}
\end{align}
where 
\begin{align}
x_{ncm}(\{r_i\},x_{cm}) = x_{cm} - \frac{m}{M_{tot}} \sum_{i=1}^{N_e} r_i.
\end{align}

The electronic and ionic kinetic energy operators transform nontrivially under our coordinate transformation.Assuming all of the ions have the same mass, $M_i = M$, the final form of $\hat{T}_n + \hat{T}_e$ after performing the change of variables is
\begin{align}
    \hat{T}_n + \hat{T}_e &= - \frac{\hbar^2}{2M_{tot}} \frac{d^2}{dx_{cm}^2} + \frac{i\hbar^2b}{M_{tot}} (N-N_e) \frac{d}{dx_{cm}} - \frac{\hbar^2}{2M}  \sum_{j,k=1}^{N-1}  \frac{d^2}{dx_i^2} + \frac{\hbar^2b^2}{2} \frac{N-N_e}{M} \nn \\ 
    &\quad- \frac{\hbar}{2NM} \sum_{j,k=1}^{N_e} \frac{d}{dr_j} \frac{d}{dr_k} + \frac{\hbar^2}{2M} \sum_{i=1}^{N_e} \frac{d^2}{dr_i^2} + \frac{\hbar^2}{2} \left( \frac{1}{M} + \frac{1}{m} \right) \sum_{i=1}^{N_e} \left( -i\frac{d}{dr_i} + b \right)^2.
\end{align}

To perform a separation of variables that removes $x_{cm}$ from the Hamiltonian, we factorize the wave function $\Psi$ as
\begin{align}
\Psi(x_{1},\ldots,x_{N-1},x_{cm},r_{1},\ldots,r_{N}) &= \chi(x_{cm}) \psi(x_{1},\ldots,x_{N-1},r_{1},\ldots,r_{N}) {,}
\end{align}
where $\chi$ obeys the normalization condition
\begin{align}
\int_{-L}^{L} dx_{cm} |\chi(x_{cm})|^2 = 1 {.}
\end{align}
Writing 
\begin{align}
\hat{\mathcal{H}}_{ring}(t) = -\frac{\hbar^2}{2M_{tot} } \frac{d^2}{dx_{cm}^2} + \frac{i\hbar^2b}{M_{tot}} (N-N_e) \frac{d}{dx_{cm}} + \hat{\mathcal{H}}(t)
\end{align}
with
\begin{align}
\hat{\mathcal{H}}(t) &= -\frac{\hbar^2}{2M} \sum_{j=1}^{N-1}  \frac{d^2}{dx_j^2} + \frac{\hbar^2}{2}\left( \frac{1}{M} + \frac{1}{m} \right) \sum_{j=1}^N \Big[ -i \frac{d}{dr_i} + b(t) \Big]^2 \nn \\
&\quad -\frac{\hbar^2}{2NM} \sum_{j,j'=1}^N \frac{d}{dr_j} \frac{d}{dr_{j'}}+\frac{\hbar^2}{2M} \sum_{j=1}^N \frac{d^2}{dr_j^2} + \frac{\hbar^2b^2(t)}{2} \frac{N-N_e}{M} + V(\hat{x}, \hat{r}) {,}
\end{align}
the Schr\"odinger equation becomes
\begin{align}
\bigg[ -\frac{\hbar^2}{2M_{tot} } \frac{d^2}{dx_{cm}^2} + \frac{i\hbar^2b}{M_{tot}} (N-N_e) \frac{d}{dx_{cm}} + \hat{\mathcal{H}}(t) \bigg] \chi \psi = \lambda_{total} \chi \psi {.}
\end{align}
We now take $\chi$ to be an eigenstate of the equation
\begin{align}
\left[-\frac{\hbar^2}{2M_{tot} } \frac{d^2}{dx_{cm}^2} + \frac{i\hbar^2b}{M_{tot}} (N-N_e) \frac{d}{dx_{cm}} \right] \chi &= \lambda_{cm} \chi.
\end{align}
Hence, the internal Schr\"odinger equation for $\psi$ becomes
\begin{align}
\hat{\mathcal{H}}(t) \psi = \lambda \psi,
\end{align}
where $\lambda = \lambda_{total} - \lambda_{cm}$.

The interparticle interactions are taken to be smooth nondivergent interactions instead of Coulomb interactions. This does not represent a limitation of the methodology as in principle calculations could be made in higher dimensions and employ Coulomb interactions without any significant modification to the methodology.  We use the potential
\begin{align}
    \hat{V}(x) = \cos^2\left[\arctan\left(z\tan\left(\frac{x}{2}\right)\right)\right] = \frac{1}{1 + z^2\tan^2\left(x/2\right)}.
\end{align}
The parameter $z$ determines the anharmonicity of the potential, whose minimum becomes narrower or flatter as $z$ moves away from $1$. 

We now switch to atomic units and set $\epsilon=1/M$, so that the Hamiltonian in the second quantization representation is
\begin{align}
    \hat{\mathcal{H}}(t) &= -\frac{\epsilon}{2} \sum_{i=1}^{N-1}  \frac{d^2}{dx_i^2} - \frac{\epsilon}{2N} \sum_{\sigma\sigma'}\sum_{k_1k_2} k_1k_2 c_{k_1\sigma}^\dagger c_{k_2\sigma'}^\dagger c_{k_2\sigma'} c_{k_1\sigma} \nn \\
    &\quad + \frac{\epsilon+1}{2} \sum_{\sigma k_1} \left[ k_1 + b(t) \right]^2 c_{k_1\sigma}^\dagger c_{k_1\sigma} + \frac{\epsilon}{2} \frac{N-1}{N} \sum_{\sigma k_1} k_1^2 c_{k_1\sigma}^\dagger c_{k_1\sigma}\nn \\
    &\quad + \frac{1}{2}\sum_{\substack{i,j=1 \\ i\neq j}}^N\left[1 + z^2\tan^2\left(\frac{1}{2}\sum_{k=1}^{N-1}x_k\left[(V^{-1})_{ki} - (V^{-1})_{kj}\right]\right)\right]^{-1} \nn \\
    &\quad + \frac{1}{2} \sum_{\sigma\sigma'} \sum_{k_1k_2q} V_q^{ee}(z) c_{k_1+q\sigma}^\dagger c_{k_2-q\sigma'}^\dagger c_{k_2\sigma'} c_{k_1\sigma} \nn \\
    &\quad - \sum_{j=1}^N \sum_{\sigma k_1 q} V_q^{ee}(z) e^{-iq\sum_{k=1}^{N-1} x_k (V^{-1})_{kj}} c_{k_1+q\sigma}^\dagger c_{k_1\sigma} + \frac{\epsilon b^2(t)}{2} (N-N_e) {.}
\end{align}
The operators $c_{k\sigma}$ are the annihilation operators of an electron with momentum quantum number $k$ and spin $\sigma$.  $V^{ee}_q(z)$ are the Fourier series coefficients of the potential. 

For numerical calculations, we construct a matrix representation of the Hamiltonian in a basis formed by the tensor product of ionic vibrational and electronic states
\begin{align}
|\nu_1\dotso\nu_{N-1}\rangle\otimes|k_1\dotso k_{N_e/2}\rangle_\uparrow\otimes|k_{N_e/2+1}\dotso k_{N_e}\rangle_\downarrow {.}
\end{align}
The $\nu$s and $k$s are harmonic oscillator and plane wave quantum numbers respectively.  In the evaluation of observables and Hamiltonian matrix elements, we use a one-to-one mapping between the many-electron basis state and an ordinal index \cite{liang1995,jia2018}.

To define harmonic oscillator basis functions adapted to our problem, we Taylor expand the effective potential in Eq.~(\ref{eq: first order adiabatic Schrodinger}) to second order.  We can set the induced gauge potential $A_{\mu,nn}$ to zero.  The resulting Hamiltonian for the $n$th eigenstate is
\begin{align}
    H_n = -\frac{\epsilon}{2} \sum_{\mu} \frac{d^2}{dx_\mu^2} + \sum_{\mu} \frac{K_{n\mu}+\epsilon K_{geo,n\mu}}{2}(x_\mu-x_\mu^0)^2.
\end{align}
As we are interested in the ground state, we set $n=0$. The harmonic oscillator basis functions we use for the matrix elements are then
\begin{equation}
    f_\nu(x) = \sqrt{\frac{(K/\epsilon)^{1/4}}{\sqrt{\pi}\nu!2^\nu}} H_\nu[(K/\epsilon)^{1/4}(x - x_0)] \text{exp}\left[-\frac{(K/\epsilon)^{1/2}(x - x_0)^2}{2}\right],
\end{equation}
where $K=K_{0\mu}+\epsilon K_{geo,0\mu}$ and $H_\nu$ are the Hermite polynomials.

Since the ions make only low amplitude oscillations with respect to the high symmetry configuration, it is reasonable to Taylor expand the ion-ion potential with respect to displacements from this configuration. We truncate the Taylor expansion to second order.  As a result, all of the Hamiltonian matrix elements can be evaluated analytically, which accelerates our computations and allows us to reach larger basis sets.  

\begin{figure}[h]
\begin{center}
        \includegraphics[width=0.4\textwidth]{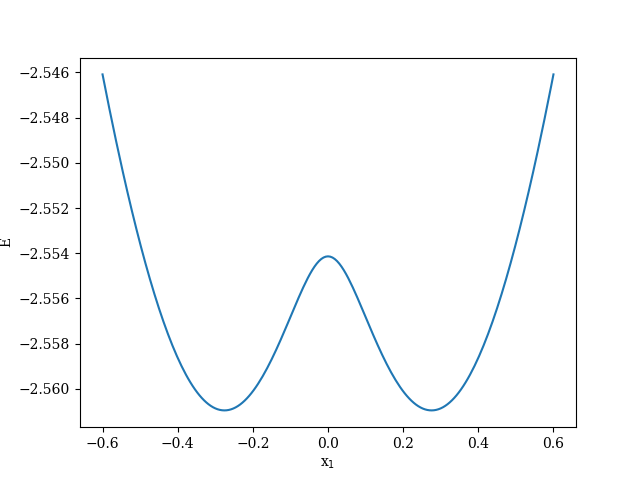}
        \includegraphics[width=0.4\textwidth]{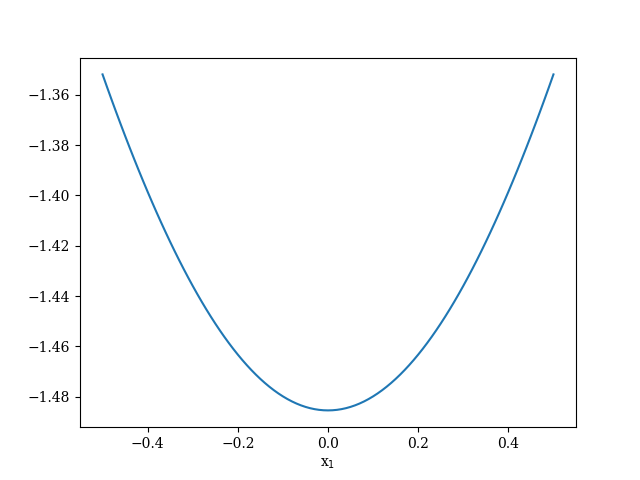}
\end{center}
    \caption{Ground state potential energy surfaces along coordinate $x_1$ for systems with $N_e=4$ and $N_e=6$ electrons.  (a) For $N_e=4$, $N=4$, and $z=1.1$ the potential energy surface has a maximum at the high symmetry point and two degenerate minima corresponding to dimerized configurations.  (b) For $N_e=6$, $N=4$, and $z=1.5$, the potential energy surface has a single minimum at the high symmetry point.  The other parameters are $k_{max}=8$ and $\epsilon=1/2000$.  \label{fig: 4 el ground PES}}
\end{figure}

We want to focus our attention first on the ideal case that the system crystallizes into the high symmetry configuration with equally spaced ions in the $\epsilon\rightarrow 0$ limit.  We have found that the system with $N_e=4$ electrons and $N=4$ ions does not adopt such a high symmetry configuration.  Figure~\ref{fig: 4 el ground PES} shows that the ground state potential energy surface has a double well along $x_1$.  The high-symmetry configuration turns out to be a local maximum of the potential, while the minima correspond to a dimerized lattice.   Instead, a system with $N_e=6$ and $N=4$ does crystallize into the high symmetry configuration as its ground state potential energy surface has a single absolute minimum as shown in right panel in Figure~\ref{fig: 4 el ground PES}.

\section{Electrical conductivity \label{sec:conductivity}}


Electrical conductivity is the linear response of charge current density to an external electric field $\mathcal{E}(x,t)$,
\begin{align}
    j_{e\alpha}(x,t) = \int_{-\infty}^t\text{d}t\int_V\text{d}x\, \sigma_{\alpha\beta}(x,t,x',t')\mathcal{E}_\beta(x',t').
\end{align}
In one dimension and for uniform fields it simplifies to
\begin{align}
    j_e(t) &= \int_{-\infty}^t\text{d}t\, \sigma(t,t')\mathcal{E}(t') \\
    j_e(\omega) &= \sigma(\omega)\mathcal{E}(\omega).
\end{align}

In an electron-only system, the current operator for electrons is given by
\begin{align}
    \hat{j}(r) = \frac{1}{2mi}\sum_\sigma \left[ \hat{\psi}^\dagger_\sigma(r) \nabla_r\hat{\psi}_\sigma(r) - [\nabla_r\hat{\psi}^\dagger_\sigma(r)] \hat{\psi}_\sigma(r)\right] - A(r,t) \hat{\psi}^\dagger_\sigma(r) \hat{\psi}_\sigma(r).
\end{align}
Since our electron-ion system has translational symmetry, we made a change of coordinates to split off the center-of-mass coordinate.  The new electronic coordinates are given by Eq.~(\ref{eq: coordinate transform}).  As a consequence, the electronic current in the new frame of reference is
\begin{align}
    \hat{j}(t) &= (1+\epsilon)\sum_{k\sigma} [k + b(t)] c_{k\sigma}^\dagger c_{k\sigma} + \epsilon b(t) (\hat{N}-\hat{N}_e) \nn \\
    &= \hat{j}_p + b(t)\hat{\tilde{N}}.
\end{align}
The first term is the paramagnetic current density operator and the second term is the diamagnetic current density operator, which has a contribution from the effective (charge) number operator $\hat{\tilde{N}}=(1+\epsilon) \sum_{\sigma k} c_{k\sigma}^\dagger c_{k\sigma} + \epsilon(N-N_e)$ when the system is not neutral.  The charge current density operator for electrons is $j_e(t) = -ej(t)$. 

\begin{figure}[h]
        \includegraphics[width=0.7\textwidth]{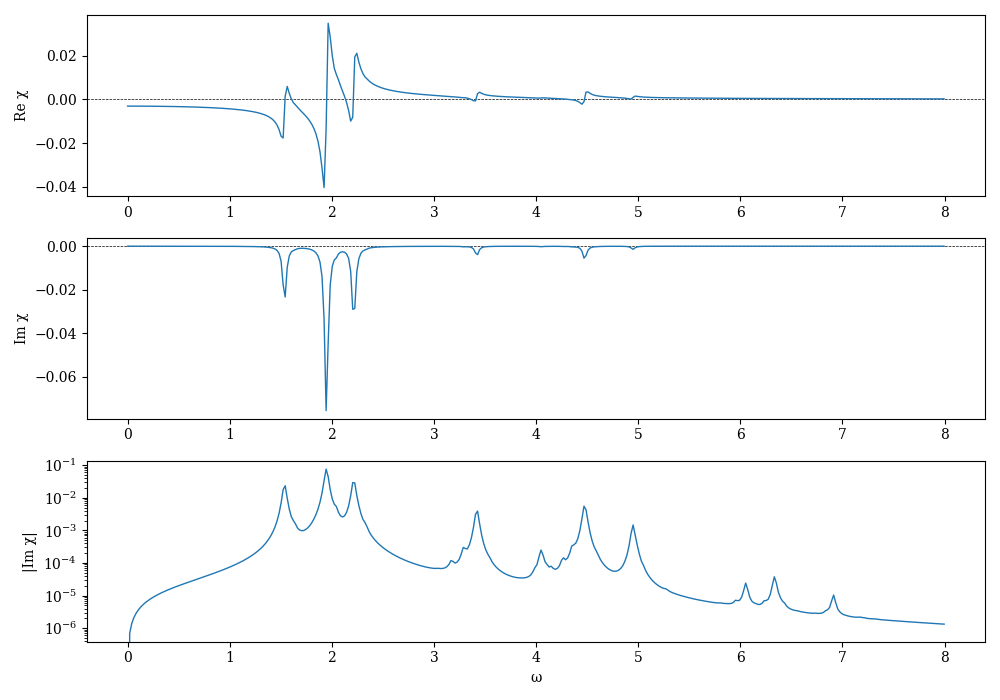}
    \caption{The frequency dependence of the response function $\chi_{j_pj_p}(\omega)$ with $\eta=0.02$ for the model of Sec.~\ref{sec:model} with 6 electrons and 4 ions and parameters $k_{max}=\nu_{max}=3$, $\epsilon=1/2000$ and $z=1.5$.}
    \label{fig: response function}
\end{figure}

\begin{figure}[t]
    \centering
    \includegraphics[width=0.7\linewidth]{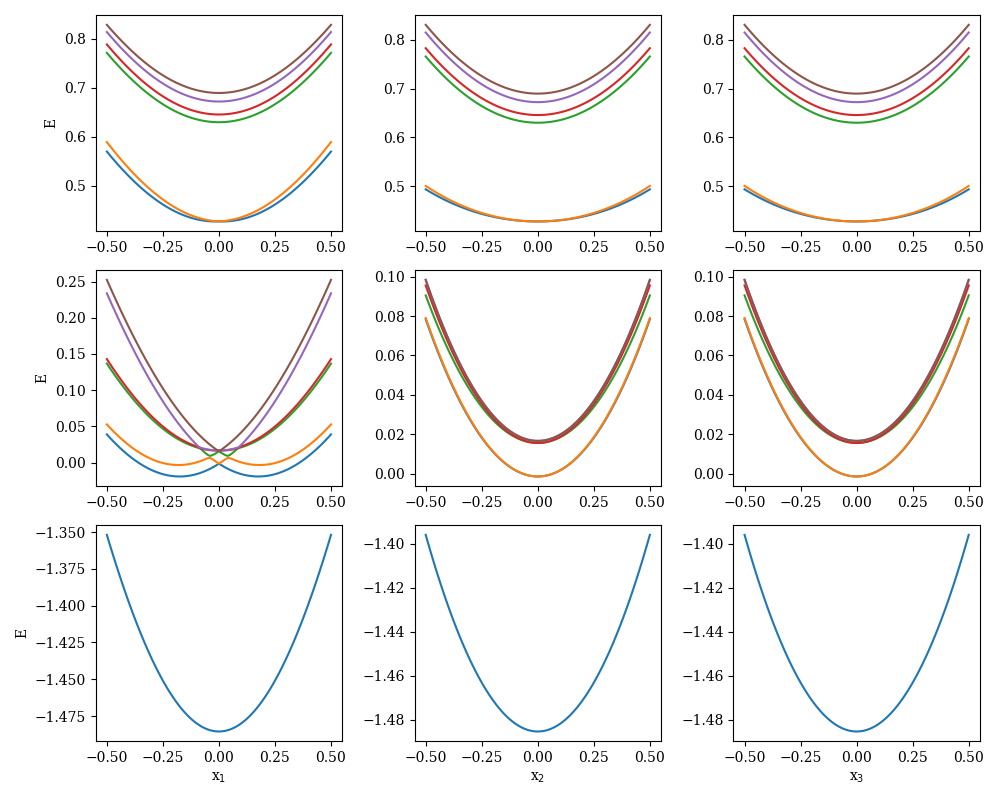}
    \caption{The first 14 potential energy surfaces for the model of Sec.~\ref{sec:model} with 6 electrons and 4 ions as functions of the ionic vibrational coordinates; parameters $k_{max}=8, \epsilon=1/2000$ and $z=1.5$. }
    \label{fig: first few PES}
\end{figure}

We want to examine the linear response of the paramagnetic part of the charge current density to a weak electric field. As shown above, in one dimension, an electric field couples to a system of electrons through a time-dependent gauge potential. The resulting linear perturbation is $b(t)j_p$. The relevant correlation function is
\begin{align}
    \chi_{j_pj_p}(t,t') = i\theta(t-t')\langle\psi_{H,0}|[ j_{p,H}(t), j_{p,H}(t') ]|\psi_{H,0}\rangle.
\end{align}
For a ground state,  the Fourier transform is 
\begin{align}
    \chi_{j_pj_p}(\omega) = \lim_{\eta\rightarrow0} \sum_{m\neq0} \left[ \frac{|\langle\psi_0| j_{p}|\psi_m\rangle|^2}{\omega - (\Lambda_m -\Lambda_0) +i\eta} - \frac{|\langle\psi_0| j_{p}|\psi_m\rangle|^2}{\omega + (\Lambda_m -\Lambda_0) +i\eta} \right].
\end{align}
The real and imaginary parts of $\chi_{j_pj_p}(\omega)$ for the model electron-ion system are shown in Fig.~\ref{fig: response function}.  Figure~\ref{fig: response function} can be compared to Fig.~\ref{fig: first few PES}, which shows the first 14 potential energy surfaces along the vibrational modes $x_1$, $x_2$, and $x_3$.  It can be seen that the first three peaks of the response function correspond to excitations from the ground state to the first three clusters of potential energy surfaces.  The other peaks correspond to clusters of high energy potential energy surfaces that are not plotted.

We can now evaluate the linear response charge current and its Fourier transform
\begin{align}
    j_e(t) &= \int_{-\infty}^\infty \chi_{j_pj_p}(t,t')b(t') \,\text{d}t' - b(t)\langle\hat{\tilde{N}}\rangle \nn \\
    j_e(\omega) &= \chi_{j_pj_p}(\omega)b(\omega) - \tilde{N}b(\omega) {.}
\end{align}
From $\mathcal{E}(t) = -\partial_t b(t)$, we get $\mathcal{E}(\omega) = i\omega b(\omega)$ and the conductivity
\begin{align}
    \sigma(\omega) = \frac{1}{i\omega} (\chi_{j_pj_p}(\omega) - \tilde{N}).
\end{align}

Because $\chi_{j_pj_p}(\omega)$ is an autocorrelation, its real part can be found by simply evaluating the limit $\eta\rightarrow0$, i.e.
\begin{align}
    \text{Re}\, \chi_{j_pj_p}(\omega) = \chi^{(1)}_{j_pj_p}(\omega) 
    &= \sum_{m\neq0} |\langle\psi_0| j_{p}|\psi_m\rangle|^2 \frac{2(E_m -E_0)}{\omega^2 - (E_m -E_0)^2} {.}
\end{align}

\begin{figure}[h]
    \centering
    \includegraphics[width=0.5\linewidth]{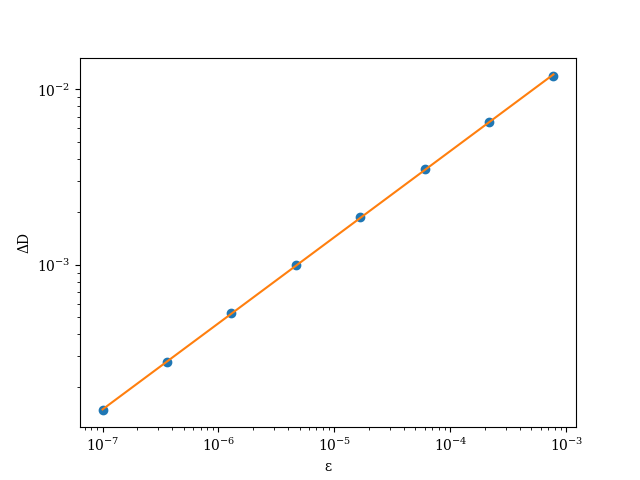}
    \caption{The difference of Drude weights for the full electron-ion system and for the electron-only system. Drude weights were calculated using Kohn's formula  for the model of Sec.~\ref{sec:model} with six electrons and four ions and parameters $k_{max}=\nu_{max}=3$ and $z=1.5$. The figure shows the error caused by neglecting the ionic degrees of freedom. The linear fit to the data is $-0.89 + 0.49\ln\epsilon$.}
    \label{fig: Drude weight difference}
\end{figure}

The Drude weight is the zero frequency part of the longitudinal conductivity and is used to distinguish insulators from metals.  In the latter, it quantifies the strength of the divergent peak in the conductivity at $\omega=0$. One of its definitions is $ D = \pi \lim_{\omega\rightarrow0} \omega \,\text{Im}\, \sigma^L(\omega)$,\cite{Resta_2018_Drude_weight} which yields
\begin{align}
    D &= \pi \lim_{\omega\rightarrow0} \omega \,\text{Im}\, \frac{1}{i\omega} (\chi^{(1)}_{j_pj_p}(\omega) - \tilde{N}) \nn \\
    &= \pi[\tilde{N}-\chi^{(1)}_{j_pj_p}(0)] {.}
    \label{eq:Drude1}
\end{align}
The evaluation of Drude weight using this formula can be costly as it involves sums over many matrix elements. A computationally easier alternative is Kohn's formula \cite{Kohn1964}
\begin{align}
    D = \pi\left.\frac{\partial^2 E_0}{\partial b^2}\right\rvert_{b=0} {,}
\end{align}
where $E_0$ is the ground state energy as a function of $b$.  As this formula only contains ground state quantities, it is much simpler to evaluate numerically. We can use it to find differences between adiabatic approximations and the exact solution. In Fig.~\ref{fig: Drude weight difference}, we plot the difference between the Drude weight calculated for the full electron-ion system and the Drude weight calculated in the Born-Oppenheimer approximation with fixed ions.  
This difference is the nonadiabatic contribution to the Drude weight, and the figure shows that it scales as $\e^{1/2}$.  This is expected because the lowest order ionic correction to the Born-Oppenheimer energy is of order $\e^{1/2}$ in the model system.

\section{Discussion}
\label{sec:discussion}

The naive application of regular perturbation theory does not yield valid zeroth order solutions of the original Schr\"odinger equation, as discussed in Sec.~\ref{sec:APT}.  Despite this, one can proceed by introducing a complete set of electron-ion basis states of the form
\begin{align}
|\psi_{n\g}(x)\rangle = \chi_{n\g}(x) |n(x)\rangle {.}
\label{eq:basis}
\end{align} 
The set of nuclear wave functions $\chi_{n\g}(x)$ indexed by $\g$ is assumed to form a complete set for each $n$.  There is a large degree of arbitrariness in the $\chi_{n\g}(x)$.  Born and Oppenheimer \cite{born1927} and Born and Huang \cite{born1954}proposed effective Schr\"odinger equations to fix the $\chi_{n\g}(x)$. We will return to this point below. 

Given a choice of nuclear functions $\chi_{n\g}(x)$, the unperturbed Hamiltonian $\mathcal{H}_0 = H\otimes I_{ion}$, where $I_{ion}$ is the identity operator on the ionic space, and the perturbation $\mathcal{H}_1 = T$ can be represented as matrices in the complete basis of states in Eq.~(\ref{eq:basis}).  The matrix elements are
\begin{align}
\mathcal{H}_{0,n\g,n'\g'} &= \delta_{nn'} \int dx \chi_{n\g}^*(x) E_n(x) \chi_{n\g'}(x) \nn \\
\mathcal{H}_{1,n\g,n'\g'} &= -\f{1}{2} g_0^{\m\n} \int dx \chi_{n\g}^*(x) \langle n(x) | \p_{\m} \p_{\n} \Big[ |n'(x)\rangle \chi_{n'\g'}(x) \Big] {,}
\label{eq:H0H1}
\end{align}
where we assume the mass tensor $g_{0\m\n}$ is $x$-independent in the present discussion.  With these ingredients, one could use conventional perturbation theory to calculate higher order corrections with respect to the perturbation $\e \mathcal{H}_1$.  Regular perturbation theory performed in this way has the following disadvantages.  First, the accuracy of the results depends on the size of the finite basis set used to represent the Hamiltonian in numerical calculations.  To achieve a given order of accuracy with respect to $\e$, the results must be converged with respect to the size of the basis set to at least that level of accuracy.  Second, each order of perturbation theory brings in an additional summation over the electronic and ionic indices $n\g$.  The effectiveness of regular perturbation theory in producing nonadiabatic corrections is dependent on the choice of basis.

In Born and Oppenheimer's perturbation theory in terms of nuclear displacements $x^{\m}-x_0^{\m}$ with respect to the reference point $x_0$, the solvability conditions require that the $\chi_{n\g}$ satisfy the harmonic effective Schr\"odinger equation 
\begin{align}
\left[ \e T + E_n(x_0) + \f{1}{2} \f{\p^2 E_n}{\p x^{\m} \p x^{\n}}\Big|_{x=x_0} (x^{\m}-x_0^{\m}) (x^{\n}-x_0^{\n})\right] \chi_{n\g} = \Lambda_{n\g} \chi_{n\g} {.}
\label{eq:BO}
\end{align}
In the special case that the state is localized, the solutions provide a complete set of basis states that are adapted to the $\e\rightarrow 0$ limit.  However, the accuracy of $\Lambda_{n\g}$ is of order $\e^{1/2}$, which is less than the order $\e$ accuracy of the zeroth order effective equation, Eq.~(\ref{eq: first order adiabatic Schrodinger}), in adiabatic perturbation theory.  Born and Oppenheimer's perturbation theory is capable of yielding systematic corrections (in powers of $\e^{1/4}$) to the zeroth order solution, provided the state is localized.

Born and Huang expressed the electron-ion state in the adiabatic representation, 
\begin{align}
|\psi(x)\rangle = \sum_n \chi_{n}(x) |n(x)\rangle{,}
\end{align}
and neglected coupling between different electronic states, to obtain [Eq.~(VIII.12) in Ref.~\onlinecite{born1954}] the equation
\begin{align}
\left[ \e T + E_n(x) - \f{\e}{2} g_0^{\m\n} \langle n |\p_{\m}\p_{\n} n\rangle \right] \chi_{n\g} = \Lambda_{n\g} \chi_{n\g} {.}
\label{eq:BH}
\end{align}
This equation is not limited to localized states, but Born and Huang gave no prescription for obtaining higher order corrections in powers of $\e$.  Additionally, the Berry-Mead-Truhlar gauge potential was set to zero in Born and Huang's equation, which is incorrect in general.  As a consequence, the potential in the last term in Eq.~(\ref{eq:BH}), called the diagonal Born-Oppenheimer correction, is not gauge- and coordinate-invariant.  The amended equation, Eq.~(\ref{eq: first order adiabatic Schrodinger}), contains the Berry-Mead-Truhlar gauge potential \cite{mead1979} and the geometric potential \cite{jackiw1988,berry1989}.  

Slater introduced the adiabatic eigenvalue problem $H|n(x)\rangle = E_n |n(x)\rangle$ and zeroth order approximation $|\psi(x)\rangle = \chi_{n\g}(x) |n(x)\rangle$ with $[\e T + E_n(x)] \chi_{n\g}=\Lambda_{n\g} \chi_{n\g}$.  Slater proposed to calculate higher order corrections by applying regular perturbation theory in the operator on the right-hand side of Eq.~(9) in Ref.~\onlinecite{slater1927}, which is different than the other approaches but has the disadvantages of regular perturbation theory mentioned above.  

The important distinctions between adiabatic perturbation theory and the approaches of Born and Oppenheimer, Slater, and Born and Huang are that adiabatic perturbation theory treats the electron-ion Schr\"odinger equation as a singularly perturbed differential equation in the $\e\rightarrow 0$ limit and produces an effective Hamiltonian for the full system of electrons and nuclei that is accurate to any order in $\e$.  Unlike regular perturbation theory, adiabatic perturbation theory does not aim at the full diagonalization of the Hamiltonian with respect to electron-ion eigenstates but rather only a partial diagonalization with respect to the electronic states.  The effective Hamiltonian that results from this procedure contains differential operators with respect to the ionic coordinates, so there is still an effective ionic Schr\"odinger equation that needs to be solved in order to obtain a full solution.

The effective Hamiltonian is useful for a variety of reasons, e.g.~it provides an alternative starting point for many-body methods.  Effective Hamiltonians often provide physical insights that help to identify the relevant interactions and develop model Hamiltonians, such as electron-phonon models, for a given system.  Since the effective Hamiltonian is Hermitian and frequency-independent, it can be straightforwardly used to evaluate thermal ensemble averages for the full electron-ion system.   Born and Oppenheimer's and Born and Huang's approaches involve two effective Hamiltonians -- one for the electrons [Eq.~(\ref{eq:adiabatic})] and one for the nuclei [Eqs.~(\ref{eq:BO}) and (\ref{eq:BH})].  Another perturbative method for the large nuclear mass limit has been proposed recently within the context of the exact factorization method \cite{tu2025}.

The version of adiabatic perturbation theory used here, which seeks the full diagonalization of the Hamiltonian with respect to the electronic states, might not be suitable for strongly nonadiabatic regimes.  In such cases, multistate variational methods may provide reasonable results \cite{perebeinos2005}.  Due to the complexity of condensed matter systems, some degrees of freedom must necessarily be neglected and this choice introduces a degree of ambiguity into variational methods.  Perturbative approaches can be applied without ambiguity, and for this reason, in problems containing small parameters and in which multiple layers of approximation are needed, there are advantages to making full use of perturbation theory before applying variational methods.  Adiabatic perturbation theory can be extended to some problems with strongly nonadiabatic coupling.  If there exists a strongly-coupled set of states which is weakly coupled to all other states, then adiabatic perturbation theory can be applied by relaxing the condition of full diagonalization to block diagonalization \cite{matyus2019,littlejohn2024}. 

Adiabatic perturbation theory has been applied to two-component quantum systems \cite{requist2025} and quantum-classical systems \cite{chatzistavrakidis2026}.  The near-identity unitary transformations in these studies are similar to Foldy and Wouthuysen's transformation of the Dirac equation to remove the coupling between positive and negative energy states \cite{foldy1950}.  The Schrieffer-Wolff transformation is also a unitary transformation that removes the coupling between two sectors of states \cite{schrieffer1966}.  There is a crucial distinction between the transformations in adiabatic perturbation theory for two-component quantum systems and Foldy-Wouthuysen and Schrieffer-Wolff transformations.  In adiabatic perturbation theory, the generators are simultaneously operators (or matrices) with respect to the electronic degrees of freedom and differential operators with respect to the ionic coordinates.  Therefore, adiabatic perturbation theory is fully representation-independent with respect to the heavy (ionic) degrees of freedom.  Instead, the generators of a Schrieffer-Wolff transformation are expressed in terms of the creation and annihilation operators of a particular basis, and are therefore tied to that choice, e.g.~the localized and conduction orbitals in the Anderson model.

\section{Conclusions}

Energy transfer and electrical transport are among the most studied properties in condensed matter systems.  Due to the high computational demands of non-empirical calculations in crystalline solids, electronic properties are often calculated for clamped nuclear positions.  This drastic approximation is reasonable in some cases yet misses important phenomena in others, such as phonon-driven superconductivity.  The predominant method for going beyond clamped nuclei involves making the harmonic approximation for the nuclei and deriving an electron-phonon model Hamiltonian.  Properties such as energy transfer and electrical transport are then usually calculated by solving the electron-phonon model with a method that makes use of the notion of quasiparticles.  

Challenges arise in time-dependent and nonequilibrium problems and in systems whose potential energy surfaces have multiple degenerate minima.  When the harmonic approximation breaks down, an adiabatic approximation might still provide an acceptable solution.  The strength of nonadiabatic coupling is an entirely different question from the validity of the harmonic approximation.

Adiabatic perturbation theory is a framework for making calculations in condensed matter systems that is conceptually different from calculations based on the harmonic approximation and quasiparticle picture.  Instead of dressing single-particle electronic and phononic propagators by the electron-phonon interaction, adiabatic perturbation theory transforms electronic or nuclear observables to a higher order adiabatic representation.  

We have found that the transformed operators have intriguing properties, e.g.~a purely electronic operator can acquire a term that contains differential operators with respect to the nuclear coordinates and the nuclear velocity operator is perturbed by a term that depends on the imaginary part of an induced Hermitian mass tensor.  The transformation could be called a dressing transformation, nevertheless the action of the transformation is mathematically distinct from, for instance, the way in which an electronic or phononic quasiparticle is dressed in many-body perturbation theory.  The transformation gives us a detailed view of the internal dynamics of how energy and charge transfer processes take place.  The generators of the unitary transformations induce virtual transitions between electronic states.

We extended adiabatic perturbation theory to time-dependent two-component quantum systems and derived several analytical results demonstrating how it is used to quantify energy and charge transfer in condensed matter systems.  We focused on two regimes -- slow dynamics and linear response.  In time-dependent problems, there are two distinct sources of nonadiabatic coupling: $d/dt$-induced coupling and nuclear kinetic energy-induced coupling. The error of approximations in adiabatic perturbation theory is controlled with respect to the small parameter $\e$. The way the error scales, e.g.~as $\e^{3/2}$ or $\e^{5/2}$, as $\e\rightarrow 0$, is known.  Results can be achieved to any desired accuracy in $\epsilon$.

To demonstrate how adiabatic perturbation theory works in practice, we introduced a one-dimensional electron-ion model system in which high accuracy numerical calculations can be performed without the need to make further approximations, such as the mean-field approximation.  Electrons and ions were fully correlated in our calculations.  We reported the electrical conductivity and the nonadiabatic correction to the Drude weight. 

\acknowledgments

R.~R. is grateful to A.~Chatzistavrakadis, L.~Jonke, and T.~Todorov for stimulating discussions.
During the completion of this work, R.~R. received support from the Croatian Science Foundation project ``Mining the Quantum: Frustration, Disorder, and Devices'' (IP-2025-02-1667) and from the European Union -- NextGenerationEU.

\bibliography{apt-energy-charge}

\end{document}